\documentclass[useAMS,usenatbib]{mn2e}
\usepackage{aas_macros,graphicx,times,multirow,amsmath,bbold}
\usepackage[]{color}
\usepackage[draft]{hyperref}

\title[Spectra and polarization of BH accretion disks]{Towards a complete description of 
spectra and polarization of black hole accretion disks: albedo profiles and returning radiation}
\author[R. Taverna et al.]{R. Taverna$^{1}$\thanks{E-mail:
\href{mailto:taverna@fis.uniroma3.it}{taverna@fis.uniroma3.it}}, 
W. Zhang$^{2}$, M. Dov\v{c}iak$^{2}$, S. Bianchi$^{1}$,
M. Bursa$^{2}$, V. Karas$^{2}$, G. Matt$^{1}$\\
$^1$Dipartimento di Matematica e Fisica, Universit\`{a} Roma Tre, via della Vasca Navale 84, I-00146 Roma, Italy\\
$^2$Astronomical Institute, Academy of Sciences of the Czech Republic, Bo\v{c}n\'{i} II 1401, CZ-14100 Prague, Czech Republic\\
}

\date{Accepted \ldots. Received \ldots; in
original form \ldots} \pagerange{\pageref{firstpage}--\pageref{lastpage}} \pubyear{2019}

\def\LaTeX{L\kern-.36em\raise.3ex\hbox{a}\kern-.15em
    T\kern-.1667em\lower.7ex\hbox{E}\kern-.125emX}

\def\der {\mathrm{d}}

\def\xims {\xi_{\rm ms}}

\def\rms {r_{\rm ms}}
\def\risco {r_{\rm ms}}
\def\rin {r_{\rm in}}
\def\rout {r_{\rm out}}

\def\loc {\mathrm{loc}}
\def\obs {\mathrm{obs}}
\def\refl {\mathrm{refl}}
\def\dir {\mathrm{dir}}
\def\ret {\mathrm{ret}}
\def\tot {\mathrm{tot}}
\def\muz {\bar{\mu}_{\rm i}}
\def\varphiz {\bar{\varphi}_{\rm i}}
\def\rg {r_{\rm g}}
\def\thetae {\bar{\theta}_{\rm e}}
\def\varphie {\bar{\varphi}_{\rm e}}
\def\thetai {\bar{\theta}_{\rm i}}
\def\varphii {\bar{\varphi}_{\rm i}}
\def\erf {{\rm erf}}

\begin{document}

\label{firstpage}
\maketitle
\begin{abstract}
Accretion disks around stellar-mass black holes (BHs) emit radiation 
peaking in the soft X-rays when the source is in the thermal state. 
The emerging photons are polarized and, for symmetry reasons, the 
polarization integrated over the source is expected to be either 
parallel or perpendicular to the (projected) disk symmetry axis, 
because of electron scattering in the disk. However, due to General 
Relativity effects photon polarization vectors will rotate with respect 
to their original orientation, by an amount depending on both the BH 
spin and the observer's inclination. Hence, X-ray polarization measurements 
may provide important information about strong gravity effects around 
these sources. 
Along with the spectral and polarization properties of radiation which
reaches directly the observer once emitted from the disk, in this paper
we also include the contribution of returning radiation, i.e. 
photons that are bent by the strong BH gravity to return again on the 
disk, where they scatter until eventually escaping to infinity. A comparison 
between our results and those obtained in previous works by different 
authors show an overall good agreement, despite the use of different 
code architectures. We finally consider the effects of absorption in 
the disk material by including more realistic albedo profiles for the 
disk surface. Our findings in this respect show that considering also 
the ionization state of the disk may deeply modify the behavior of polarization 
observables.
\end{abstract}
\begin{keywords}
stars: black holes -- relativistic processes -- accretion discs -- X-rays: binaries
-- polarization.
\end{keywords}

\section{Introduction}
\label{intro}
Black holes (BHs) are the most compact astrophysical objects, which
can be used as ideal laboratories to test physics in the presence 
of ultra-strong graviational fields. Different kinds of BHs have been 
identified so far and classified according to their mass. Stellar-mass 
BHs ($M\approx 10\,M_\odot$) are believed to be born mainly in the 
gravitational core-collapse of stars with a initial mass $\ga 25$--$30
\,M_\odot$ \cite[see][for a review]{whw02}. On the other hand, the different 
observational manifestations of active galactic nuclei (AGN) have been 
explained with the presence of supermassive BHs \cite[$M\approx$~$10^6
$--$10^9\,M_\odot$, see e.g.][]{salp64,zn64,lb69}, which have been observed 
to be hosted in centers of most galaxies, including our own Milky Way 
\cite[see][]{kh13,gr16}. Finally, intermediate-mass BHs ($M\approx10^2$--$10^5
\,M_\odot$) have been also reported in some dim AGN or associated to 
ultra-luminous X-ray sources \cite[see][for a recent review]{mez17}. 

Despite the fact that even light cannot escape from their event 
horizon, BHs have been observed so far as X-ray sources, through 
the electromagnetic radiation emitted from accretion disks in binary 
systems, as well as, in the case of intermediate and supermassive 
BHs, by studying the orbits of nearby objects (stars) influenced 
by the presence of their gravitational field \cite[as in the case 
of the supermassive BH at the Galactic center, see][]{sch+02}. 
Very recently, BHs have been in the spotlight thanks to the results 
achieved by the LIGO/VIRGO collaboration, which detected for the 
first time the gravitational waves emitted in BH-BH merging events 
\cite[][]{abb+16}. Nevertheless, electromagnetic observations still 
play a key role in understanding the physics of BHs.

General relativity plays a fundamental role in the physical processes 
that occur in the BH environment. The trajectories of photons which 
propagate close to a BH, for example, deviate from straight lines, 
following null geodesics that can be described in the proper space-time 
metric \cite[][]{bar70}. Moreover, photon frequencies turn out to be 
gravitationally red-shifted as a result of the BH potential well, whereas
the Doppler effect changes the photon frequency either way depending 
on the state of motion of the source with respect to the observer (Bardeen,
Press \& Teukolsky \citeyear{bpt72}). However, strong gravity also 
influences the polarization states of photons. In fact, the polarization 
vectors are generally rotated, with respect to their original orientation, 
because of the parallel transport along null geodesics. The recent 
developments in X-ray polarimetry techniques \cite[][]{cos+01,bell+10} 
gave indeed new impetus to this field of research.

Theoretical models to investigate spectra and polarization from these 
sources have been developed by many authors in the past \cite[see e.g.][and 
\citealt{sk13} for a comprehensive list of further references]{laor+90,
matt+93,rb94,bao+97,ak00,lnm05,lnm09,kar06}. Furthermore, spectral and polarimetric
studies on radiation coming from accretion disks and atmospheres in the
case of AGN (either magnetized or not) have been carried out by \citet[see
also \citealt{sil02,sil+18}]{sil+09}. In the present paper we revisit 
in particular the work by \citet{dov+08}, 
first incorporating in their original treatment the contribution of ``returning 
radiation'', i.e. photons which are forced by the strong BH gravity to 
return to the disk before reaching the observer at infinity. This issue 
has been already addressed by \citet[see also \citealt{sk10}]{sk09}, who 
exploited a Monte Carlo code \cite[exhaustively described in][see also 
\citealt{s+13}]{sk13} to properly take into account both vacuum transport 
and radiative processes that influence photon propagation around stellar-mass 
BHs. While we adopt Kerr metric in the present paper, we will also refer 
to the study carried out by \citet{kraw12}, who instead developed a ray-tracing 
code capable to give predictions for different kinds of space-time metrics. 

Although a complete, self-consistent treatment of ionization is still 
deferred to a future work, we move a step towards the inclusion in the 
model of a realistic ioniziation profile of the disk material. To this 
aim we convolve the output of our code with a non-trivial prescription 
for the disk albedo, obtained using the external code {\sc cloudy} \cite[][]{cloudy}. 
The results show that both spectral and polarization properties of BH 
accretion disk emission are considerably modified by considering absorption 
in the disk with respect to the simple case of $100\%$ albedo. For this 
reason, in order to correctly interpret the physical and geometrical 
information that can be extracted from these sources using X-ray polarimetry, 
a proper study of the optical properties of the disk is required. This 
acquires even more significance in the light of the launch of the NASA-SMEX 
mission {\em IXPE} \cite[][]{weiss+13}, scheduled for 2021, which will
be capable of probing X-ray polarimetric properties of bright accreting
black hole binaries.

The plan of the paper is as follows. In section \ref{section:themodel} 
we briefly set up the theoretical model and present our main assumptions, 
while an overview of the numerical implementation is described in section 
\ref{section:numericalimplementation}. Results are then illustrated in 
section \ref{section:results}: in particular, the outputs of our codes 
are compared with the results previously obtained by \citet{sk09} and 
\citet{kraw12} in \S \ref{subsection:100albedo}; the effect of changing 
the optical depth of the electron-dominated atmosphere assumed to cover 
the disk surface are discussed in \S \ref{subsection:differenttau}; and 
the results obtained considering a different albedo prescription for the 
disk are presented in \S \ref{subsection:albedo}. Finally, we summarize
our findings and present our conclusions in section \ref{section:discussion}.

\section{The model}
\label{section:themodel}
We consider in this work only accretion disks around stellar-mass BHs
in their thermal state, which are observed to emit radiation peaking
in the soft X-rays \cite[at about $1$ keV, see e.g.][]{sk09}. The space-time 
around the central BH is described by the Kerr metric, for different 
values of its dimensionless angular momentum per unit mass $a$. 

We treat the accretion disk as a geometrically-thin standard 
disk \cite[][]{ss73}. In particular, in order to dissipate their 
angular momentum, particles on the disk are assumed to rotate around 
the central BH with a Keplerian velocity. Photons are assumed to be 
emitted from the disk according to an isotropic blackbody (BB) distribution
at different temperatures, following a Novikov-Thorne radial profile 
\cite[a multicolor disk, see][see also \citealt{cunn75} for a more 
complete relativistic treatment]{nt73,wang00},
\begin{flalign} \label{equation:ntprofile}
T(M,r,a) &= 741 f_{\rm col}\left(\frac{M}{M_\odot}\right)^{-1/2}
\left(\frac{\dot{M}}{M_\odot/{\rm yr}}\right)^{1/4} \nonumber & \\
\ &\ \ \ \ \times\big[f(\xi,a)\big]^{1/4}\,{\rm keV}\,, & 
\end{flalign}
where $M$ is the BH mass, $\dot{M}$ is the mass accretion rate, $\xi=
(r/r_{\rm g})^{1/2}$, with $r$ the radial coordinate and $r_{\rm g}=GM/
c^2$ the gravitational radius, and \cite[see][]{pt74}
\begin{flalign}
f(\xi,a) &= \frac{1}{\xi^4(\xi^3-3\xi+2a)}\left[\xi-\xims-\frac{3}{2}a\ln\left(\frac{\xi}{\xims}\right)\right. \nonumber & \\
\ &\ \ -\frac{3(\xi_1-a)^2}{\xi_1(\xi_1-\xi_2)(\xi_1-\xi_3)}\ln\left(\frac{\xi-\xi_1}{\xims-\xi_1}\right) \nonumber & \\
\ &\ \ -\frac{3(\xi_2-a)^2}{\xi_2(\xi_2-\xi_1)(\xi_2-\xi_3)}\ln\left(\frac{\xi-\xi_2}{\xims-\xi_2}\right) \nonumber & \\
\ &\ \ \left.-\frac{3(\xi_3-a)^2}{\xi_3(\xi_3-\xi_1)(\xi_3-\xi_2)}\ln\left(\frac{\xi-\xi_3}{\xims-\xi_3}\right)\right]\,, &
\end{flalign}
with $\xi_{1,2}=2\cos[(\arccos{a}\mp\pi)/3]$, $\xi_3=-2
\cos[(\arccos{a})/3]$ and $\xims=(\risco/r_{\rm g})^{1/2}$,
where $r_{\rm ms}$ is the radius of the innermost stable circular
orbit \cite[ISCO, see][]{bpt72}\footnote{In the Kerr metric 
and for prograde orbits, $\risco=r_{\rm g}$ for a maximally rotating 
BH ($a\rightarrow1$) while $\risco=6r_{\rm g}$ for a non-rotating 
BH (Schwarzschild case, $a=0$).}. The hardening factor $f_{\rm col}$ 
is adopted to shift the energy of the thermal photons emerging from 
the disk in order to account (in a simplified way) for the effects 
of scatterings they undergo with disk particles \cite[see][]{dov+08,dea19}.
We also assumed no-torque at the inner edge of the disk, which is 
taken as coincident with the ISCO.


Photons emitted from the disk are expected to be linearly polarized
essentially due to scatterings that occur onto the particles which 
compose the disk. In this respect, we assume that the photon polarization 
states follow the Chandrasekhar profile \cite[see e.g.][]{chan60}; 
this amounts to consider a pure-electron, scattering dominated atmospheric 
layer that covers the disk surface. Symmetry considerations lead 
to two possible polarization states, with the polarization vector 
either parallel or perpendicular to the projection of the disk symmetry 
axis in the plane of the sky. In particular, in the hypothesis of 
no absorption in the atmosphere, polarization is perpendicular to 
the axis if the optical depth is large ($\tau>1$), while it is parallel 
to the axis for smaller values of $\tau$ \cite[see e.g.][and references 
therein]{dov+08}.

Photons propagating in the space-time around a BH experience general 
relativistic effects. More in detail, photons emitted close to a BH 
are redshifted due to the strong gravitational field, and their 
trajectory follows null geodesics, deviating from straight lines. 
As a consequence, the paths of photons emitted from regions of the 
disk closer to the central BH can be bent by its gravitational field. 
These photons may still reach the observer at infinity (direct radiation), 
but part of them can return back to the disk surface and interact with 
the disk material before eventually arriving at infinity (returning 
radiation); this depends in general on the emission point on the disk 
and on the emission direction.

In addition to the photon energy and trajectory, strong gravity can also 
influence the polarization state of radiation. In fact, the photon 
polarization plane should be parallely transported along the curved 
trajectory, and this results in a rotation of the polarization vector 
with respect to the direction assumed at the emission \cite[][]{connst77,
stconn77,cps80}. For these reasons, the expected polarization fraction
$P$ and polarization angle $\chi$ will be modified with respect to
those at the emission. In particular, the polarization angle change
$\Psi$ is given by,
\begin{flalign} \label{equation:Psi}
\tan\Psi &= \frac{Y}{X}\,, &
\end{flalign}
that can be expressed in terms of the components of the Walker-Penrose
constant $\kappa_1$ and $\kappa_2$ \cite[][see also \citealt{dov+08}
for complete expressions]{wp70,chan83} as
\begin{flalign} \label{equation:XYPsi}
X &= -(\alpha-a\sin\theta_{\obs})\kappa_1-\beta\kappa_2 \nonumber & \\
Y &= (\alpha-a\sin\theta_{\obs})\kappa_2-\beta\kappa_1\,. &
\end{flalign}
In equations (\ref{equation:XYPsi}), $\alpha$ and $\beta$ are the
impact parameters which identify the $x$ and $y$ axes, respectively,
of the observer's sky reference frame in the plane perpendicular
to the line-of-sight, and $\theta_{\obs}$ is the observer's inclination. 
Equations (\ref{equation:XYPsi}) also show that a further rotation 
\cite[called the gravitational Faraday rotation, see][]{ishi+88}, 
occurs only if the central BH is rotating with specific angular 
momentum $a\neq0$.

Polarization of photons which return to the disk can be also 
influenced by the interactions they undergo onto disk particles. 
While previous works reduced their investigation to the simplifying 
assumption of pure-scattering atmospheres with $100\%$ albedo 
at the disk surface, \cite[see e.g.][]{dov+08,sk09,kraw12}, here 
we discuss more realistic cases, in which the absorption opacity 
of the disk surface layer is not zero, exploring different albedo 
prescriptions. We finally discuss the effects of these new assumptions 
on the spectral and polarization properties of the emerging radiation.



\section{Numerical implementation}
\label{section:numericalimplementation}
\begin{figure*}
\begin{center}
\includegraphics[width=17.5cm]{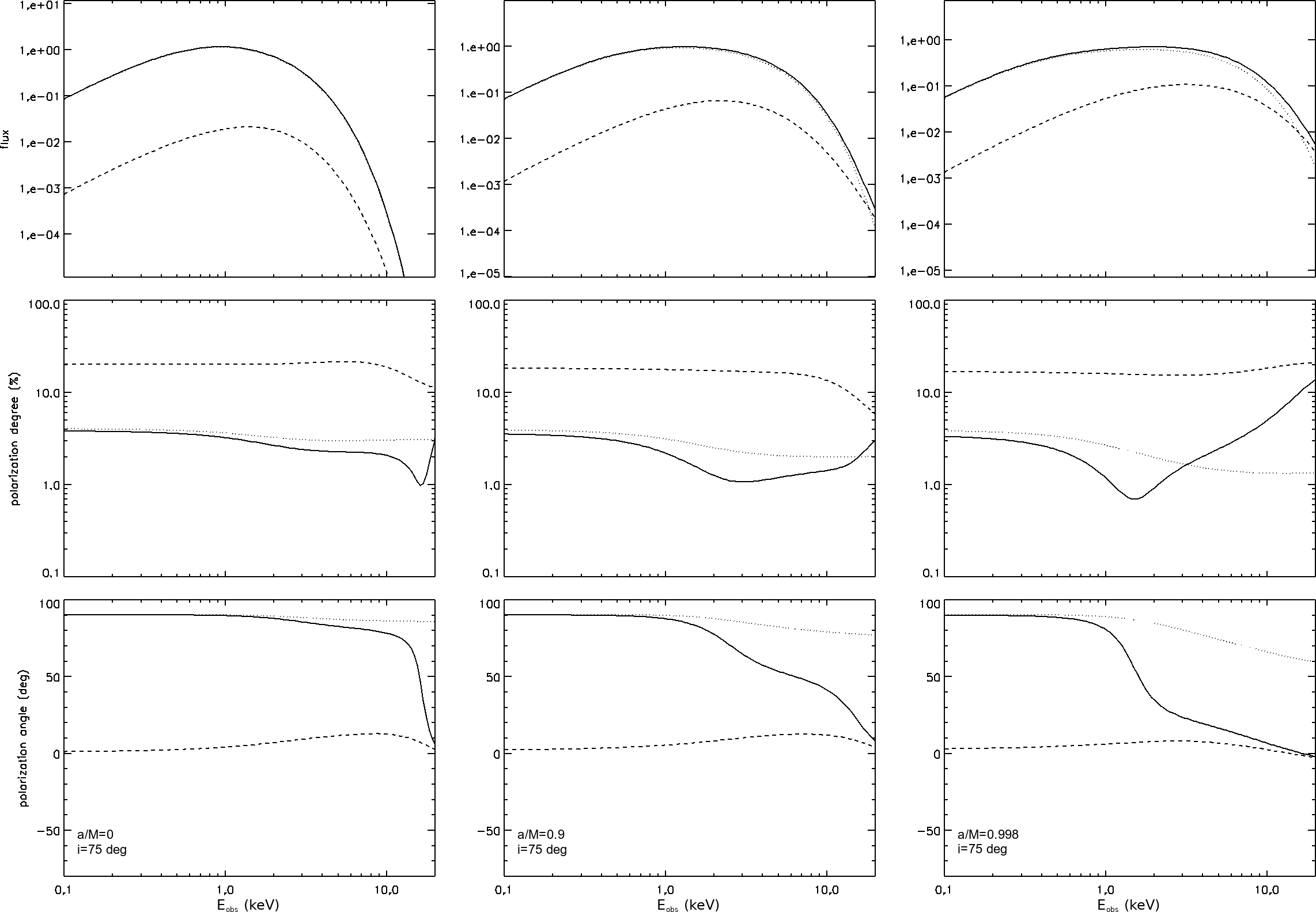}
\caption{Spectrum (top row), polarization degree
(middle row) and polarization angle (bottom row)
plotted as functions of the photon energy at the
observer for a BH mass $M=10\,M_\odot$ and spin 
$a=0$ (left-hand column), $0.9$ (middle column) 
and $0.998$ (right-hand column). The inclination 
angle $i$ between the observer's line-of-sight 
and the disk symmetry axis is taken as $75^\circ$,
while the BH mass accretion rate is chosen in
such a way that the accretion luminosity is equal 
to $10\%$ of the Eddington limit. In particular 
$\dot{M}=2.45$ (for $a=0$), $0.90$ (for $a=0.9$) 
and $0.35$ (for $a=0.998$) in units of $10^{18}$ 
g s$^{-1}$ (see Table 1 of \citealt{kraw12}). In 
each plot, the contributions of direct and returning 
photons alone are marked by dotted and dashed lines, 
respectively, while the joint contribution (direct 
$+$ returning radiation) is marked by solid lines.}
\label{figure:i75}
\end{center}
\end{figure*}
\begin{figure*}
\begin{center}
\includegraphics[width=17.5cm]{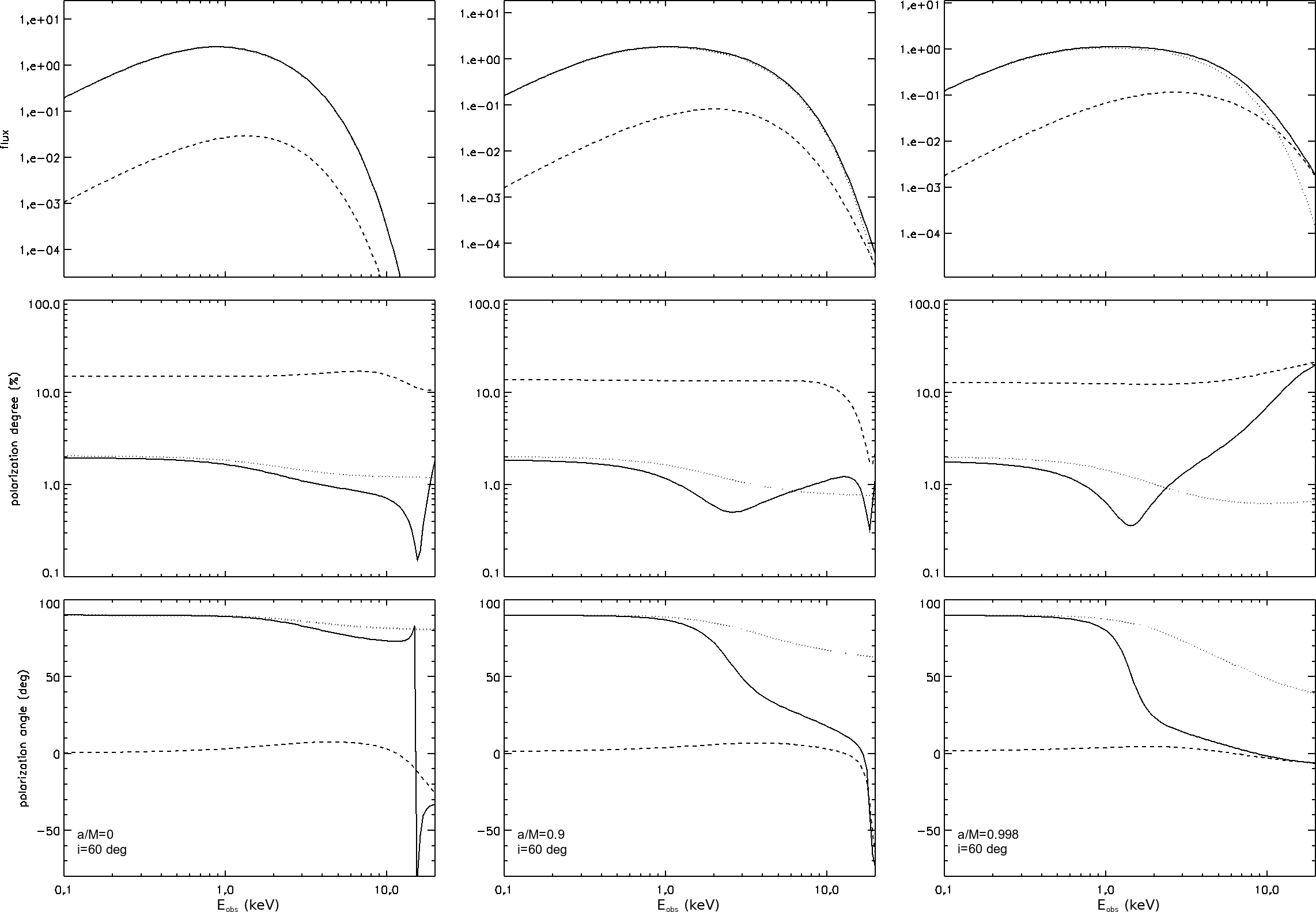}
\caption{Same as in Figure \ref{figure:i75}
but for $i=60^\circ$.}
\label{figure:i60}
\end{center}
\end{figure*}
\begin{figure*}
\begin{center}
\includegraphics[width=17.5cm]{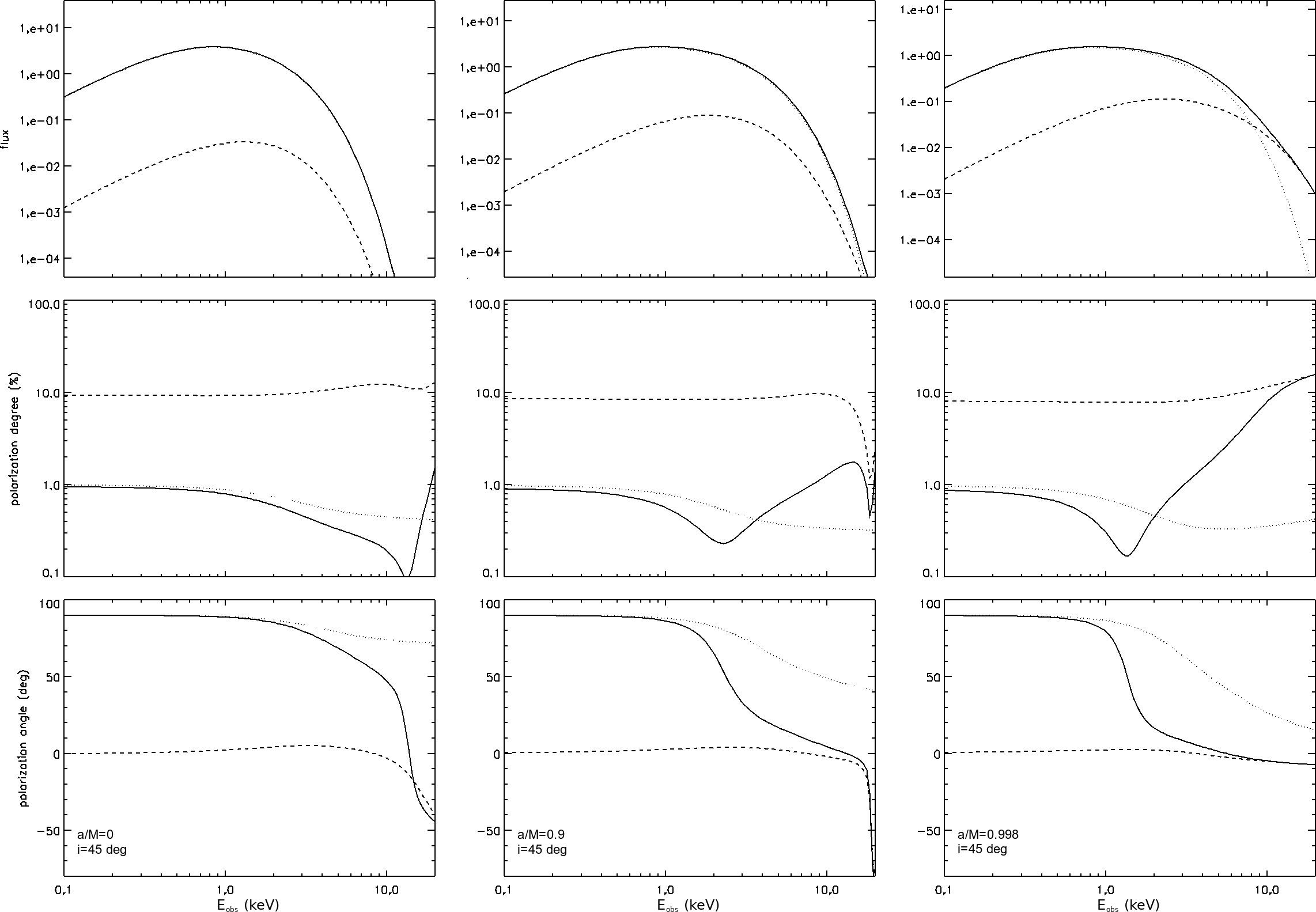}
\caption{Same as in Figure \ref{figure:i75}
but for $i=45^\circ$.}
\label{figure:i45}
\end{center}
\end{figure*}
For our calculations we used the {\sc kyn} package \cite[][]{dov04,dov+04a,dov+04b,dov+04c},
which includes a set of different emission models built-in 
into {\sc xspec}. The code exploits a fully relativistic, 
ray-tracing technique based on an observer-to-emitter 
approach. As photons emitted from the disk are assumed 
to follow a (multicolor) blackbody distribution, we resort 
to the model {\sc kynbb}, which provides the spectral and 
polarization properties of radiation in the case of blackbody 
seed photons, following a Novikov-Thorne temperature profile 
with color correction (see section \ref{section:themodel}). 
However, since in its original version this model did not 
account for returning radiation, the first part of the present 
work consisted in expanding {\sc kyn}, adding a specific 
routine to include the contributions of returning photons
to spectra and polarization observables.

In order to briefly summarize how the code works, we 
remind here in broad terms the main features of {\sc 
kyn} in dealing with the direct radiation case \cite[we 
address the reader to][for a more in-depth view]{dov+08}, 
before entering into details about the returning radiation 
section.

\subsection{Direct radiation} \label{subsec:direct}
The disk surface is sampled by a $(r,\phi)$ grid 
with $N_{\rm r}\times N_\phi$ points, where $r$ is 
the radial distance from the central BH and $\phi$ 
the azimuth with respect to a reference direction 
in the plane perpendicular to the disk axis. Once 
the observer inclination $\theta_{\obs}$ with respect 
to the disk normal is determined, the code traces 
back all the possible null geodesics which connect 
the observer to different points on the disk, followed
by direct photons.  

For each emission point of the disk surface grid, 
all the main quantities concerning the radiative 
transport are then provided in terms of photon 
number. In this respect, the photon flux $\Delta 
f_{\obs}$ observed at infinity in the energy bin 
$\langle E,E+\Delta E\rangle$ per unit solid angle 
can be written as
\begin{flalign} \label{equation:Deltafo}
\Delta f_{\obs} &= \int_{\rin}^{\rout} r\der r\int_{\phi'}^{\phi'+\Delta\phi'}
\der\phi\int_{E/g}^{(E+\Delta E)/g}Gf_{\loc}\der E_{\loc}\,, &
\end{flalign}
where the subscript `loc' refers to local quantities, 
$\rin$ ($\rout$) is the inner (outer) radius of the 
disk and $f_\loc$ is the local, energy-dependent 
photon number flux \cite[see][for further details]{dov+08}. 
The boundaries $\phi'$ and $\Delta\phi'$ of the azimuthal 
integration domain can be arbitrarily chosen in input 
(for integration over the entire disk surface it is 
$\phi'=0$ and $\Delta\phi'=2\pi$), as well as the values 
$E_{\loc}^{\rm min}$ and $E_{\loc}^{\rm max}$ which 
mark the range of variation of the local energy. Also 
$\rin$ and $\rout$ can be chosen as code inputs; in 
particular, the inner edge of the disk can be selected 
as coincident, larger or smaller than the innermost 
stable circular orbit. In the latter case, particles 
in the inner part of the disk are considered in free-fall 
towards the central BH, with the same energy and angular 
momentum they had at the ISCO. In the following, 
however, we assume $\rin=\rms$.

As mentioned above, emitted photons are assumed to 
be polarized by electron scatterings which occur at 
the disk surface. In particular, in the diffusion limit 
($\tau\gg 1$) the intrinsic polarization properties 
are given following the analytical expressions developed 
by \citet{chan60} for a geometrically-thin, optically-thick 
atmosphere with infinite optical depth. On the other 
hand, we also exploit the capability of the {\sc kyn} 
package to explore configurations characterized by 
finite optical depths. In these cases, the intrinsic 
polarization pattern is numerically evaluated using 
the Monte Carlo code {\sc stokes}, firstly developed 
by \citet[see also \citealt{mar18} and references therein 
for subsequent updates]{gg07}. It has been possible 
to verify a posteriori that all the configurations 
with $\tau\ga 5$ shall be regarded as equivalent to 
the $\tau=\infty$ case \cite[see Figure \ref{figure:varitau};
see also][]{dov+08}.


The code then calculates the local, energy-dependent 
Stokes parameters\footnote{Since the polarization of 
photons which arise from scatterings onto atmospheric 
electrons is essentially linear, we do not consider in 
our following calculations the Stokes parameter $v$,
which accounts for circular polarization.} 
$\boldsymbol{s}_{\loc}^\dir=[i_\loc^\dir,q_\loc^\dir,
u_\loc^\dir]$ and, once selected the polar angles $\theta$ 
and $\varphi$ the photon emission direction makes with 
the disk normal, it computes the integrated (energy-dependent) 
Stokes parameters at the observer as
\begin{flalign} \label{equation:integratedstokes}
i_{\obs}&=\int{\rm d}S\,i_{\loc}^\dir(r,\phi) G & \nonumber \\
q_{\obs}&=\int{\rm d}S\,[q_{\loc}^\dir(r,\phi)\cos(2\Psi)
-u_{\loc}^\dir(r,\phi)\sin(2\Psi)]G & \nonumber \\
u_{\obs}&=\int{\rm d}S\,[u_{\loc}^\dir(r,\phi)\cos(2\Psi)
+q_{\loc}^\dir(r,\phi)\sin(2\Psi)]G\,. &
\end{flalign}
In equations (\ref{equation:integratedstokes}), $\Psi$ 
is the change in polarization angle (see equation 
\ref{equation:Psi}) which accounts for the general 
relativistic effects that the photon polarization 
plane experiences as they propagate along each geodesic
which connects the emission points to the observer. 
On the other hand, the transfer function $G=g^2l\cos
\theta$, where $g=E_\obs/E_\loc$ is the energy shift 
and $l$ the lensing factor (i.e. the ratio between 
the flux tube cross sections at the observer and at 
the disk), accounts for the effects of strong gravity 
on the photon energies and directions along the same 
trajectories \cite[see][for more details]{dov04,dov+08}. 
Finally, $\der S=r{\rm d}r{\rm d}\phi$ represents the 
surface integration element. As for equation (\ref{equation:Deltafo}), 
integration is extended over the radial range $\rin<r
<\rout$ and the azimuthal range $\phi'<\phi<\phi'+\Delta
\phi'$. Polarization observables, i.e. the polarization 
degree $P$ and angle $\chi$, are eventually obtained as
\begin{flalign} \label{equation:polarizationobservables}
P &= \frac{\sqrt{q^2+u^2}}{i} & \nonumber \\
\chi &= \frac{1}{2}\arctan\left(\frac{u}{q}\right)\,, &
\end{flalign}
where $i$, $q$ and $u$ should be substituted with the
expressions given in equations (\ref{equation:integratedstokes}).

\subsection{Returning radiation} \label{subsec:returning}
In order to include the contributions of returning radiation
inside {\sc kyn}, we use the {\sc c++} code {\sc selfirr}, 
based on the ray-tracing {\sc sim5} package \cite[][]{bur17},
that computes all the possible null geodesics connecting
two different points on the disk surface, along which 
returning photons move. 

Returning photons are considered to collide with the disk 
along a general direction $(\bar{\theta}_{\rm i},\bar{\varphi}
_{\rm i})$, with $\bar{\theta}_{\rm i}$ and $\bar{\varphi}_{\rm i}$ 
the polar angles with respect to the disk normal. The disk 
surface is therefore divided into $\bar{N}_{\rm r}$ incidence
patches, each characterized by the radial distance $\bar{r}_{\rm i}$ 
of their centers from the BH, while all the possible incidence 
directions are in turn sampled through a discrete $\bar{N}_\theta
\times\bar{N}_\varphi$ angular mesh. An orthonormal tetrad $e_{(a)}^\mu$
is then attached to the disk fluid at each radius, following 
Appendix A.2 of \citet{zdb19}; in this way, in the tetrad frame
comoving with the disk fluid, the wave vector associated to
each returning photon can be written as
\begin{flalign}
k^{(a)}&= \{1, \sin\thetai\cos\varphii, \cos\thetai, \sin\thetai\sin\varphii\} &
\end{flalign}
or, by switching to the Boyer-Lindquist coordinate frame, as
\begin{flalign}
k^\mu&= e^\mu_{(a)} k^{(a)}\,. &
\end{flalign}
Following \citet{car68} and \citet{lnm05}, the code exploits 
$k^\mu$ to evaluate the constants of motion which characterize 
each photon and determine whether the photon trajectory actually 
connects the incidence point to another point of the disk surface
or not. If yes, the code solves the geodesic equation for the 
radius $\bar{r}_{\rm e}$ of the point at which the photon was 
emitted from the disk and calculate the angles $\thetae$ and 
$\varphie$ which identify the emission direction with respect 
to the disk normal \cite[see][]{lnm05}\footnote{We stress that 
{\sc selfirr} naturally accounts for photons that, due to lensing, 
can travel along more than one geodesic between two points on 
the disk surface with the same direction of incidence.}. This 
also allows to compute the value of the energy shift 
\begin{flalign}
\bar{g}&=\frac{E_{\rm i}}{E_{\rm e}}=\frac{k_{\mu{\rm i}}U^\mu_{\rm i}}{k_{\mu{\rm e}}U^\mu_{\rm e}} &
\end{flalign}
which accounts for general relativistic effects along each returning 
geodesic, with $U_{\rm i}^\mu$ ($U_{\rm e}^\mu$) the four-velocity
of the disk fluid at the incidence (emission) point\footnote{We remind
that the disk material is considered to rotate around the central BH
with Keplerian velocity.}. The angle $\bar{\Psi}(\bar{r}_{\rm i},
\bar{\theta}_{\rm i},\bar{\varphi}_{\rm i})$ by which the photon polarization 
plane is rotated while returning to the disk is eventually provided. 
This is done again by evaluating the Walker-Penrose constant (as mentioned 
in section \ref{section:themodel}), following equations B46--B47 of 
\citet{lnm09}.

The code returns in output a fits file to be read by {\sc kyn}. 
This contains, for each value of $\bar{r}_{\rm i}$, the values 
of the incidence angles $(\bar{\theta}_{\rm i},\bar{\varphi}_
{\rm i})$, the radial distance $\bar{r}_{\rm e}$ of the starting 
point and the emission angles $(\bar{\theta}_{\rm e},\bar{\varphi}
_{\rm e})$, the energy shift $\bar{g}$, the solid angle $\Delta\bar{\Omega}_{\rm 
i}$ of the pixel with incidence angles $\bar{\theta}_{\rm i}$ and 
$\bar{\varphi}_{\rm i}$ and the change in polarization angle $\bar{\Psi}$. 
Input parameters of each {\sc selfirr} run are, instead, the specific 
BH spin $a$, the disk outer radius\footnote{The inner radius is set 
as coincident to $\rms$ following the choice of $a$.} $\rout$ and 
the number of points $\bar{N}_{\rm r}$, $\bar{N}_\theta$ and $\bar{N}_\varphi$ 
of the different discrete grids introduced above. For the sake 
of convenience, we chose $\rout$ as coincident to the disk outer 
radius set in the calculations of the direct radiation contributions 
(see section \S \ref{subsec:direct}), much in the same way as 
the numbers of radial grid points $\bar{N}_{\rm r}$ and $N_{\rm 
r}$. This ensures that the radial grids which sample the disk 
surface in the two codes {\sc kyn} and {\sc selfirr} are the 
same, so that the contributions of direct and returning radiation 
can be simply summed together at each point of this unique surface 
grid without any additional numerical manipulation (see equation 
\ref{equation:sumdirret}).

As in the case of the direct radiation, the Stokes parameters 
$\boldsymbol{\bar{s}}_{\rm e}=[\bar{i}_{\rm e},\bar{q}_
{\rm e},\bar{u}_{\rm e}]$ of returning photons at their 
emission are given following the Chandrasekhar's (\citeyear{chan60})
formulae or, alternatively, using the tables obtained
from the  Monte Carlo code {\sc stokes} (see \S \ref{subsec:direct}), 
according to their emission radius $\bar{r}_{\rm e}$ and 
direction $(\bar{\theta}_{\rm e},\bar{\varphi}_{\rm e})$. 
Returning photons are then reflected at the disk surface.
In order to evaluate the Stokes vectors $\boldsymbol{\bar{s}}
_\refl(P_\refl,\chi_\refl)=[\bar{i}_\refl,\bar{q}_\refl,
\bar{u}_\refl]$, where $P_\refl$ and $\chi_\refl$ denote
the polarization degree and angle after reflection, we 
use the Chandrasekhar's (\citeyear{chan60}) formulae for 
diffuse reflection. First, the code computes the Stokes 
parameters for three distinct states of polarization, 
corresponding to unpolarized light ($P_\refl=0$) and fully 
polarized radiation ($P_\refl=1$) with $\chi_\refl=0$ 
and $45^\circ$, respectively (see Appendix \ref{appendix:drformulae}
for more details). Then it reconstructs the Stokes vector 
for a generic state of polarization through the decomposition
\begin{flalign} \label{equation:combinationstokesvector}
\boldsymbol{\bar{s}}_\refl(P_\refl,&\chi_\refl)=\boldsymbol{\bar{s}}_\refl(0,-) \nonumber & \\
\ &\,\,\,\,\,+P_{\rm e}
\{[\boldsymbol{\bar{s}}_\refl(1,0)-\boldsymbol{\bar{s}}_\refl(0,-)]
\cos[2(\chi_{\rm e}+\bar{\Psi})] \nonumber & \\
\ &\,\,\,\,\,+[\boldsymbol{\bar{s}}_\refl(1,\pi/4)-\boldsymbol{\bar{s}}_
\refl(0,-)]\sin[2(\chi_{\rm e}+\bar{\Psi})]\}\,, &
\end{flalign}
which follows from the definition (\ref{equation:polarizationobservables})
of $P$ and $\chi$ in terms of the Stokes parameters, with 
$P_{\rm e}$ and $\chi_{\rm e}$ the polarization degree 
and angle at the emission point.


All the contributions coming from the different incidence 
directions at each incidence point are finally summed together, 
obtaining the contribution to the (energy-dependent) local 
Stokes parameters for returning photons,
\begin{flalign} \label{equation:localstokesret}
i_\loc^\ret(\bar{r}_{\rm i})&=\sum_{\bar{\theta}_{\rm i},\bar{\varphi}_{\rm i}}
\bar{i}_\refl(\bar{r}_{\rm i},\bar{\theta}_{\rm i},\bar{\varphi}_{\rm i})
\bar{g}^2(\bar{r}_{\rm i},\bar{\theta}_{\rm i},\bar{\varphi}_{\rm i})\bar{\mu}_{\rm i}\Delta\bar{\Omega}_{\rm i}
\nonumber & \\
q_\loc^\ret(\bar{r}_{\rm i})&=\sum_{\bar{\theta}_{\rm i},\bar{\varphi}_{\rm i}}
\bar{q}_\refl(\bar{r}_{\rm i},\bar{\theta}_{\rm i},\bar{\varphi}_{\rm i})
\bar{g}^2(\bar{r}_{\rm i},\bar{\theta}_{\rm i},\bar{\varphi}_{\rm i})\bar{\mu}_{\rm i}\Delta\bar{\Omega}_{\rm i}
\nonumber & \\
u_\loc^\ret(\bar{r}_{\rm i})&=\sum_{\bar{\theta}_{\rm i},\bar{\varphi}_{\rm i}}
\bar{u}_\refl(\bar{r}_{\rm i},\bar{\theta}_{\rm i},\bar{\varphi}_{\rm i})
\bar{g}^2(\bar{r}_{\rm i},\bar{\theta}_{\rm i},\bar{\varphi}_{\rm i})\bar{\mu}_{\rm i}\Delta\bar{\Omega}_{\rm i}\,, 
&
\end{flalign}
where $\bar{\mu}_{\rm i}=\cos{\bar{\theta}_{\rm i}}$.
The integrated Stokes parameters at the observer are 
computed following the same procedure described in 
section \ref{subsec:direct} for direct radiation; 
in particular, equations (\ref{equation:integratedstokes})
become,
\begin{flalign} \label{equation:integratedstokestotal}
i_{\obs}&=\int{\rm d}S\,i_{\loc}(r,\phi) G & \nonumber \\
q_{\obs}&=\int{\rm d}S\,[q_{\loc}(r,\phi)\cos(2\Psi)
-u_{\loc}(r,\phi)\sin(2\Psi)]G & \nonumber \\
u_{\obs}&=\int{\rm d}S\,[u_{\loc}(r,\phi)\cos(2\Psi)
+q_{\loc}(r,\phi)\sin(2\Psi)]G\,, &
\end{flalign}
where, in this case\footnote{As pointed out above, 
we remind here that the Stokes parameters of direct 
and returning photons are defined over the same radial
grid.},
\begin{flalign} \label{equation:sumdirret}
i_{\loc}(r,\phi)&=i_\loc^\dir(r,\phi)+i_\loc^\ret(r) \nonumber & \\
q_{\loc}(r,\phi)&=q_\loc^\dir(r,\phi)+q_\loc^\ret(r) \nonumber & \\
u_{\loc}(r,\phi)&=u_\loc^\dir(r,\phi)+u_\loc^\ret(r)\,. &
\end{flalign}

\begin{figure*}
\begin{center}
\includegraphics[width=11.7cm]{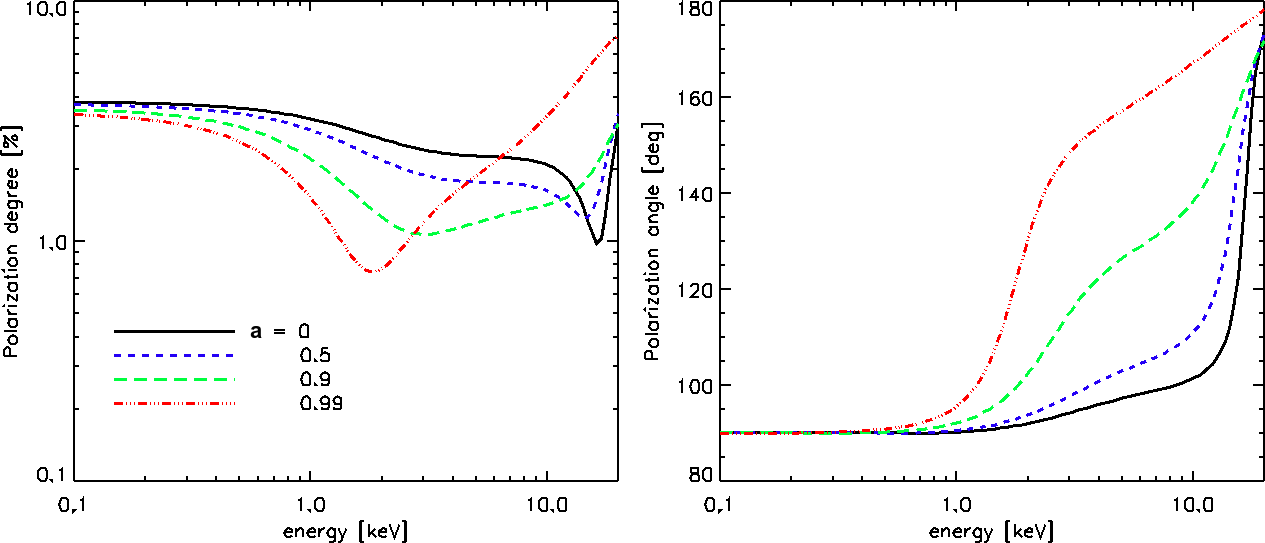}
\caption{Polarization degree (left) and polarization 
angle\textsuperscript{\ref{krawwarn}} (right), plotted 
as functions of the photon energy at the observer, 
for $i=75^\circ$ and different values of the BH spin: 
$a=0$ (solid black), $0.5$ (dashed blue), $0.9$ (long-dashed 
green) and $0.99$ (dash-dotted red). Input parameters 
are taken as in \citet[see his Table 1]{kraw12}.}
\label{figure:krawplot}
\end{center}
\end{figure*}
The output of a typical {\sc kyn} run consists in a table
containing the values of the integrated Stokes parameters 
$\boldsymbol{s}_{\obs}=[i_{\obs},q_{\obs},u_{\obs}]$ 
as functions of the photon energy. This also allows to
obtain the energy-dependent polarization observables $P_{\obs}$ 
and $\chi_{\obs}$, by substituting equations 
(\ref{equation:integratedstokestotal}) into equations 
(\ref{equation:polarizationobservables}). Alternatively, 
spectra and polarization properties can be also obtained 
for either direct or returning radiation contributions 
separately, by omitting in equations (\ref{equation:sumdirret})
the returning or the direct terms, respectively.

\section{Results} \label{section:results}
With the purpose of an immediate comparison of our 
results with previous works \cite[in particular][]{sk09,kraw12}, 
and in order to ensure that our code is well implemented, 
we initially assume a 100\% albedo prescription, 
i.e. all the photons which return to the disk 
are considered to be reflected towards infinity. 
As an extension of the above-mentioned works, 
we also obtain results for finite optical depths 
of the pure-electron disk atmosphere. Finally, 
we discuss how the spectral and polarization 
properties of the collected radiation can be 
modified assuming more realistic albedo profiles. 

\subsection{Case of 100\% disk albedo} \label{subsection:100albedo}
We firstly remark that, contrary to our ray-tracing 
code, which is based on an observer-to-emitter 
approach, those exploited by \citet{sk09} and 
\citet{kraw12}, to which we refer in the following
as benchmarks, rely on an emitter-to-observer paradigm
\cite[see][for more details]{sk13}. In the light of this, 
a comparison with their results becomes even more 
significant, since it also allows to check if our 
code gives similar expectations to those already 
presented in literature and obtained using different 
techniques.

In order to make our simulations comparable with 
those presented by both \citet{sk09} and \citet{kraw12}, 
we assumed a central BH with mass $M=10\,M_\odot$ 
and accreting material at a rate $\dot{M}\approx10\%$ 
of the Eddington limit $\dot{M}_{\rm Edd}$ (correctly
accounting for the change in efficiency for different 
values of the BH spin $a$). Moreover, we chose a
constant hardening factor $f_{\rm col}=1.8$. The energy 
domain is characterized by a 200-point grid between $0.01$ 
and $50$ keV, while the disk surface has been 
sampled through a grid characterized by $N_{\rm r}=500$ 
radial bins, logaritmically-spaced between $\rms$ 
and $\rout=100\,\rg$, and $N_\phi=180$ azimuthal 
zones. We then associated to each point $\bar{N}
_\theta\times\bar{N}_\phi=100\times100$ incidence 
directions for what concerns returning radiation, 
i.e. $10^4$ rays are traced per radius and azimuth\footnote{We 
checked a posteriori that increasing the resolution 
of the incidence direction grid at more than $100
\times100$ points does not modify significantly 
the output.}. Finally, the polarization properties 
of photons at their emission are derived using 
Chandrasekhar's (\citeyear{chan60}) formulae, 
considering an infinite optical depth $\tau$ 
for the atmospheric layer that is assumed to 
cover the disk surface.

Figures \ref{figure:i75}--\ref{figure:i45} show 
the behaviors of spectra (energy flux of the 
Stokes parameter $i$) and polarization observables 
($P_\obs$ and $\chi_\obs$) as functions of the 
photon energy at the observer, considering the 
line-of-sight inclined by $75^\circ$, $60^\circ$
and $45^\circ$ with respect to the disk symmetry 
axis and for three different values of the BH 
spin: $a=0$, $0.9$ and $0.998$. In all the plots 
the contributions of direct and returning radiation 
alone (dotted and dashed lines, respectively) 
can be distinguished from the joint contribution
(solid lines) obtained by summing the two, as
illustrated in \S \ref{subsec:returning}. 

\begin{figure*}
\begin{center}
\includegraphics[width=17.5cm]{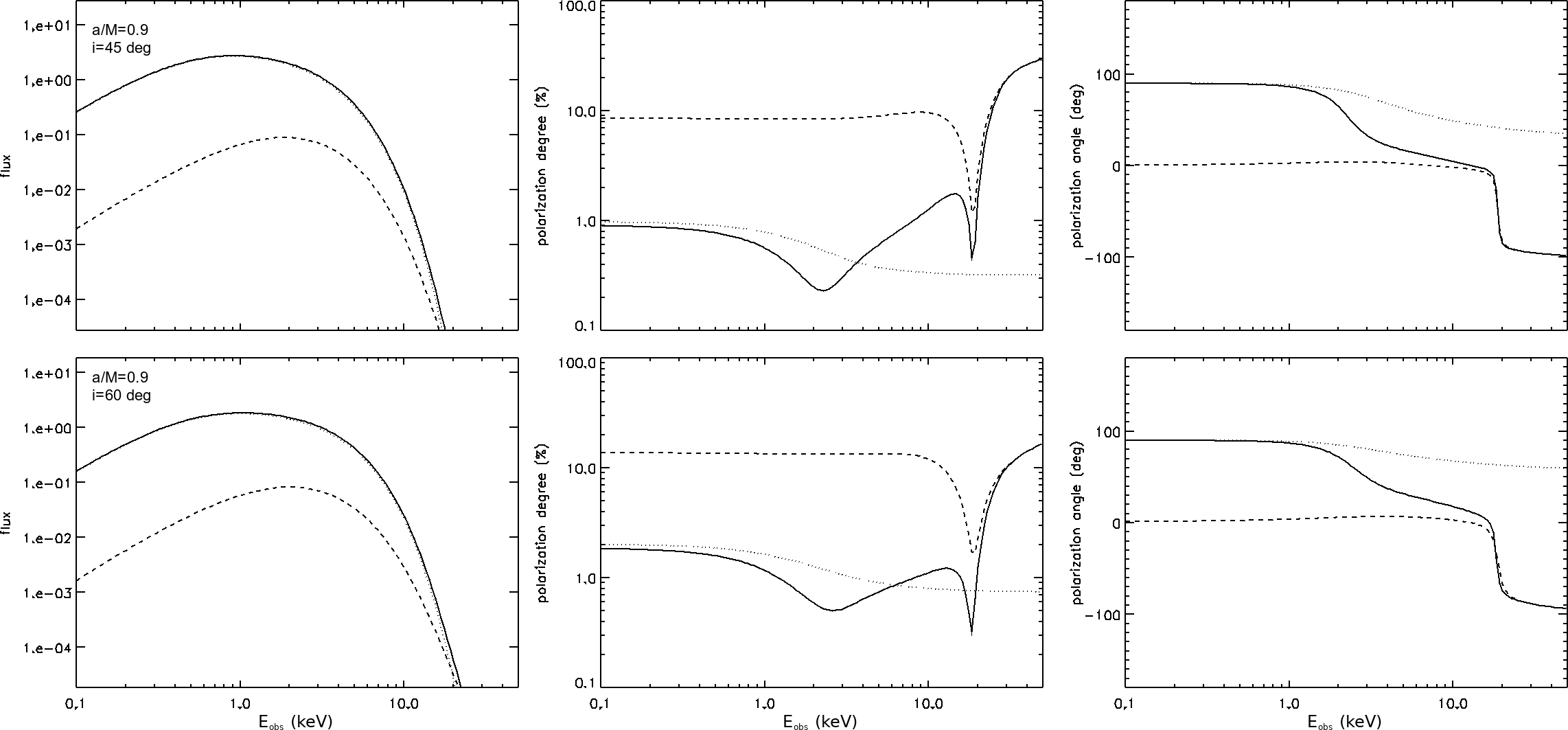}
\caption{Spectrum (left-hand column), polarization degree 
(middle column) and polarization angle (right-hand column),
plotted as functions of the photon energy at the observer,
for $a=0.9$ and $i=45^\circ$ (top row) and $i=60^\circ$ 
(bottom row). Details as in Figures \ref{figure:i75}--\ref{figure:i45}.}
\label{figure:feature}
\end{center}
\end{figure*}
As a quick comparison with the plots reported 
in their Figures 4--6 clearly shows\footnote{We 
warn the reader that, at variance with \citet{sk09}, 
in {\sc kyn} a polarization angle of $0^\circ$ 
is associated to polarization parallel to 
the disk symmetry axis \cite[see][]{dov+08}.},
our results turn out to be in general compatible
with those presented by \citet{sk09}. In particular, 
as expected for large optical depths \cite[see e.g.][]{dov+08}, 
direct radiation turns out to be polarized perpendicularly 
to the disk symmetry axis at lower energies, i.e.
the polarization angle associated to the direct 
radiation component $\chi^\dir_\obs$ is about $90^
\circ$, while it slowly decreases at higher energies 
($\ga 2$ keV) under the effect of the polarization 
plane rotation (see section \ref{section:themodel}). 
This can be explained looking at the behavior 
of the disk temperature profile as a function 
of the radial distance from the center (see 
e.g. Figure \ref{figure:albedovsEvsR}). Low-energy 
photons are emitted from the external regions 
of the disk, where the temperature is lower 
and strong gravity is less important, whereas 
high-energy ones are emitted closer to the BH, 
where the temperature is higher and photon
polarization is mostly affected by general
relativistic effects. On the other hand, 
returning photons appear to be mostly polarized 
parallel to the disk axis ($\chi^\ret_\obs=
0^\circ$), although also in this case the 
polarization angle slightly declines above 
$10$ keV due to general relativistic effects 
(even if by a smaller amount than in the direct 
radiation case). Looking at the joint contribution 
(direct $+$ returning radiation), a transition 
between the two regimes can be observed. In 
particular, the total polarization angle $\chi
_\obs$ follows the curve of direct radiation 
alone as long as the fraction of returning 
photons becomes comparable to that of direct 
ones (see the spectra in the top rows), while
it swings by $90^\circ$ at higher frequencies.
The energy at which this transition occurs 
turns out to be smaller the larger the 
BH spin, that is mainly due to the choice
of considering $\rin$ as coincident to
$\rms$. In fact, the innermost stable 
circular orbit lies at a larger distance 
from the center for a non-rotating BH, while 
it approaches the horizon as $a$ increases. 
As a consequence, for $a=0$ returning 
photons start to dominate only at very high
energies ($\ga 10$ keV), once the spectrum 
of direct photons has sufficiently declined, 
while for $a=0.9$ and $0.998$ the transition 
occurs already at $E_\obs\ga1$ or $2$ keV, since 
more high-energy returning photons populate 
the spectral tail at those energies.

The trend of the polarization degree can
be explained in a rather similar way. Having
assumed that the radiative transfer in the 
atmospheric layer which covers the disk surface 
is dominated by (Thomson) electron scatterings 
with $\tau\rightarrow\infty$, direct radiation 
turns out to be quite mildly polarized, with 
$0.3\%\leq P^\dir_\obs\leq3\%$. Returning 
radiation, instead, is much more polarized, 
with in general $8\%\leq P^\ret_\obs\leq20\%$. 
As in the polarization angle plots, a transition 
between the direct and returning radiation regimes 
can be observed when contributions are summed
together. While the $P^{\rm tot}_\obs$ essentially 
follows the behavior of direct radiation below 
$\sim 1$ keV, it attains a minimum in correspondence
of the polarization angle swing described
above, and then it approaches the returning
radiation trend at higher energies.

For the sake of a further comparison with 
works previously published in literature,
we also produced supplementary plots (see 
Figure \ref{figure:krawplot}) for the energy-dependent 
polarization degree and angle, to be compared
with those reported by \citet{kraw12}, who 
showed expectations for different kinds of 
space-time metrics. Also in this case, limiting 
our analysis to the case of a Kerr BH (see his 
Figure 5\footnote{Since in \citet{kraw12}
the polarization angle is assumed to be $0$
for polarization parallel to the disk plane,
increasing for a clockwise rotation of the
polarization plane, we reorganized the output
of {\sc kyn} in Figure \ref{figure:krawplot}
with the purpose of a better comparison.\label{krawwarn}}), 
an overall good agreement should be noted between 
the different simulations performed for the 
polarization observables behaviors, obtained 
for $a=0$, $0.5$, $0.9$ and $0.99$ and $i=75
^\circ$.

\begin{figure*}
\begin{center}
\includegraphics[width=17.5cm]{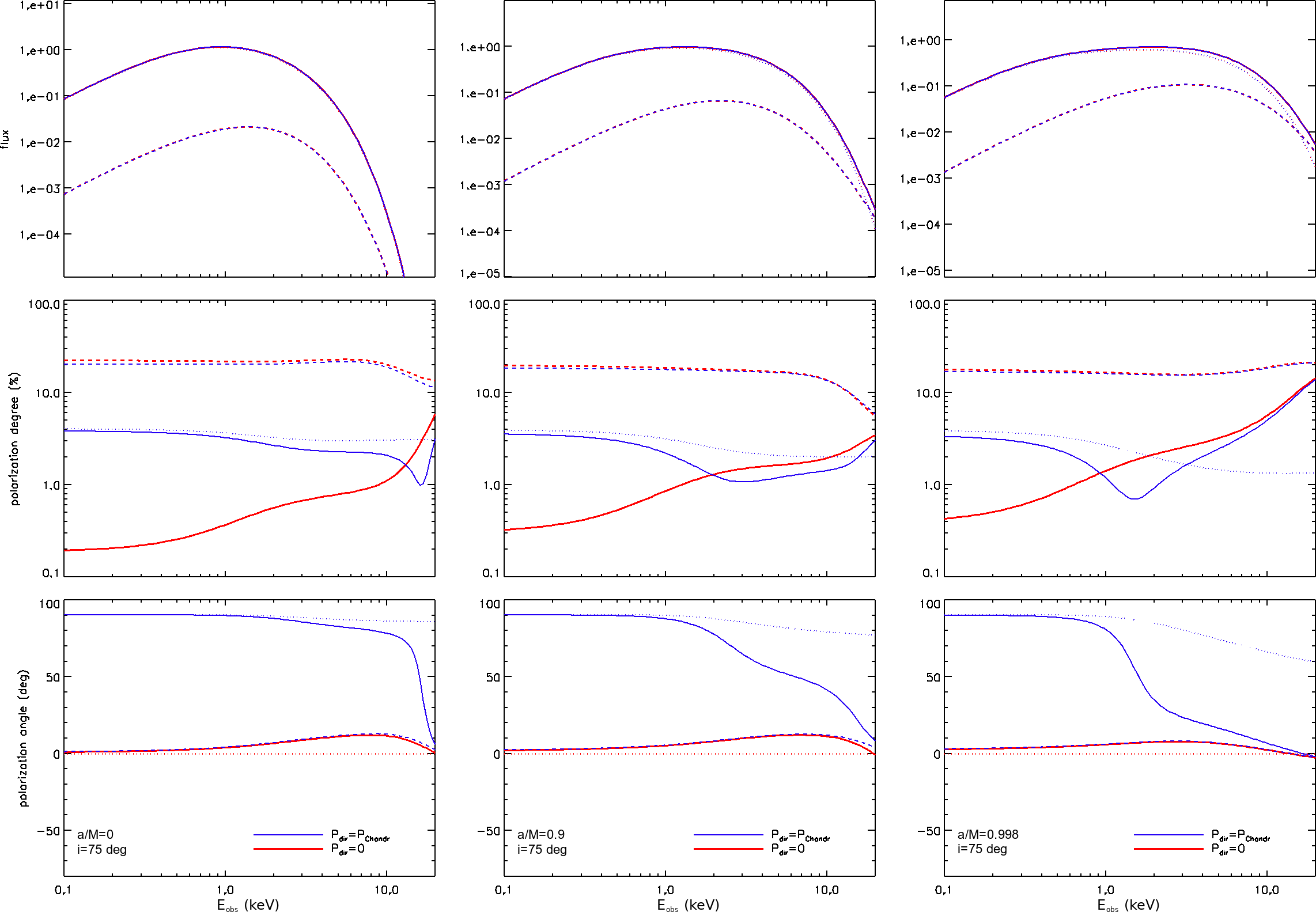}
\caption{Spectrum (top row), polarization degree (middle
row) and polarization angle (bottom row) behaviors,
plotted as functions of the photon energy at the observer,
for direct radiation polarized according to Chandrasekhar's
(\citeyear{chan60}) formulae ($P_\dir=P_{\rm Chandr}$, 
blue lines) and assuming that emitted radiation is unpolarized 
($P_\dir=0$, red lines). Parameter values and line styles
are taken as in Figure \ref{figure:i75} (see also text for 
more details).}
\label{figure:directoff}
\end{center}
\end{figure*}
Some differences with respect to the results
reported in \citet{sk09}, however, are present
when the case of $a=0.9$ is considered. This 
is already visible in the middle column of Figures 
\ref{figure:i60} and \ref{figure:i45}, where 
a second minimum in the (joint) polarization 
degree behavior appears in addition to that 
just discussed above, related to the transition 
from the direct to the returning radiation 
regimes. This second minimum occurs in connection
with a steep decrease in the polarization 
angle of the returning radiation component,
at an energy $\sim 20$ keV. To better investigate 
this feature, we report in Figure \ref{figure:feature} 
the same plots (for $a=0.9$ and $i=45^\circ,\,60^\circ$), 
but extending the energy range on the horizontal 
axis up to $50$ keV. As it can be clearly seen in 
the right panels, for these values of the input 
parameters the polarization direction of returning 
photons seems change from parallel ($\chi_{\rm obs}^
{\rm ret}=0^\circ$) to perpendicular ($\chi_{\rm 
obs}^{\rm ret}=-90^\circ$) to the disk symmetry 
axis at very high energies. 
We checked a posteriori that this behavior is 
still present also increasing the resolution 
of the energy, radial and angular grids; this 
led us to the conclusion that this particular 
feature may not be a numerical artifact. 

In order to understand the physical mechanism
responsible for this behavior, we looked in more 
details on the values of the polarization degree
and angle of the reflected radiation as functions
of the position on the disk surface. 
We found that an additional critical point 
\cite[as discussed in][]{dov+08,dov+11} appears close to 
the BH for the largest spin values and the highest 
inclinations, around which the polarization angle 
attains all possible values. The existence of this 
critical point is due to the complex dependence of 
Chandrasekhar's (\citeyear{chan60}) diffuse reflection 
formulae (in the multi-scattering approximation) on 
the incidence and emission scattering angles, as 
well as on the photon polarization angle at the 
reflection point.
In most cases, the rather small region around the 
critical point does not affect sizeably the overall 
Stokes parameter distributions. High-energy photons come from 
a well-defined, ``hot'' region of the disk close to
the ISCO where the temperature is higher. The Doppler 
shift has to be important, which means that it is on 
the approaching side of the disk, but at the same time
the gravitational redshift is not too large, which means
that it is not too close to the BH horizon. If the critical 
point lies far away from this peculiar region, one does not see any 
abrupt change in the polarization properties at high 
energies, since the Stokes parameters do not change 
too much in the vicinity of this zone. The situation 
is different when the critical point is close to this 
hot region. In fact, while the Stokes parameters at low-energies 
are integrated over a large area of the disk surface,
those at high energies come from a localized region
and this results in different polarization patterns.
In particular, in our case it turns out that the critical 
point is close enough to the hot region only for not too high
spin values ($a=0.9$). For $a\simeq0$ the critical point does not even exist (i.e. 
it would lie below the ISCO), while it moves closer to 
the horizon, and on the other side of the BH wrt the hot 
region, for very high spins ($a=0.998$).

However, as it can be noted from the spectra reported 
in the left panels of Figure \ref{figure:feature}, 
at such high energies ($\ga 20$ keV) the photon 
flux is dramatically smaller than at the spectral 
peak (by almost $5$ orders of magnitude). This
certainly reduces the overall importance of 
catching such a behavior in the polarization 
observable trends, since polarimetric techniques 
(especially at X-ray energies) would suffer 
for the extremely low number of photons
expected.


\begin{figure*}
\begin{center}
\includegraphics[width=17.5cm]{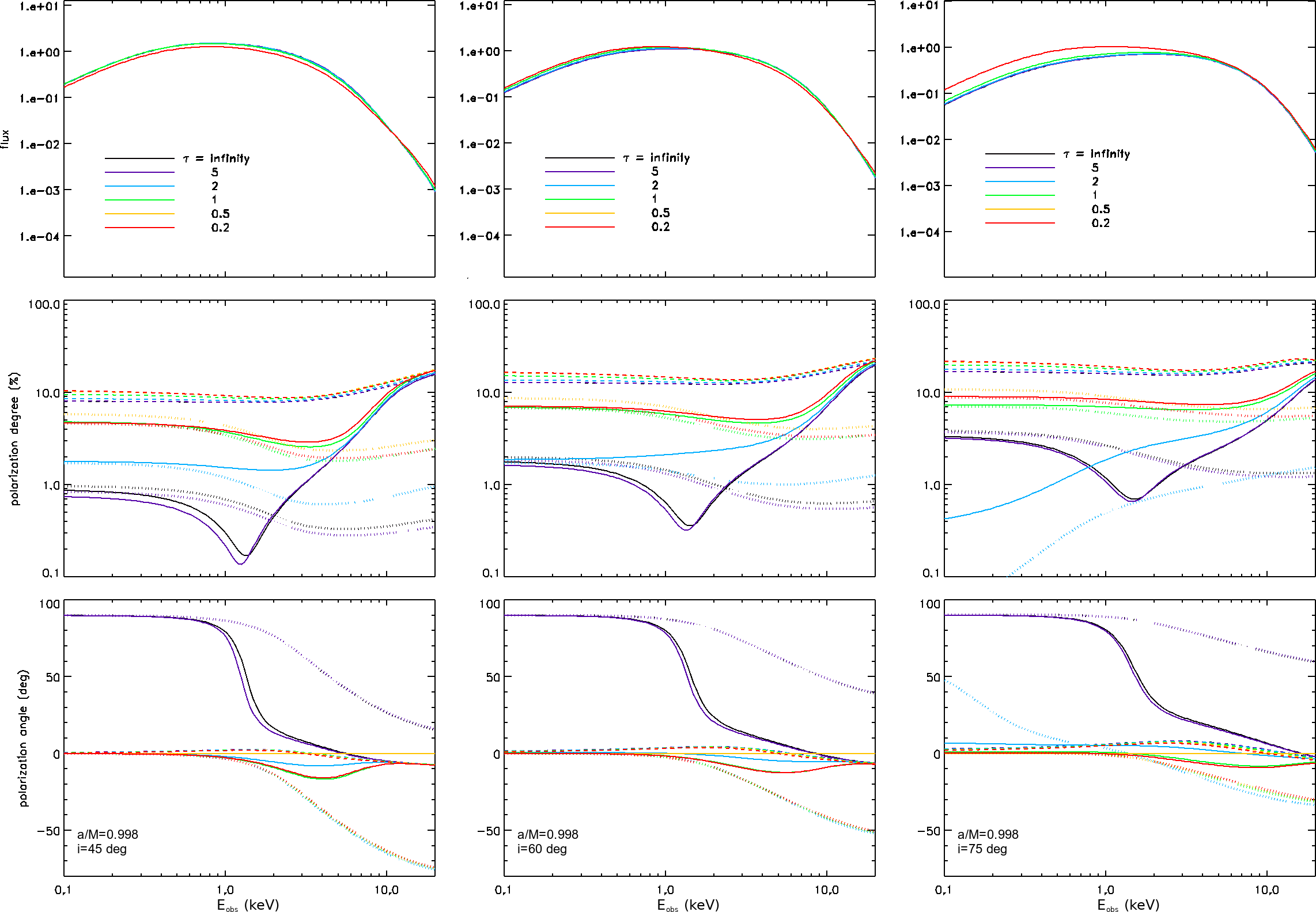}
\caption{Spectrum (top row), polarization degree 
(middle row) and polarization angle (bottom row) 
plotted as functions of the photon energy at the
observer for $a=0.998$ and $i=45^\circ$ (left-hand 
column), $60^\circ$ (middle column) and $75^\circ$ 
(right-hand column), and for different value of the 
atmosphere optical depth: $\tau\rightarrow\infty$ 
(black), $5$ (purple), $2$ (cyan), $1$ (green), 
$0.5$ (orange) and $0.2$ (red). Details as in
Figure \ref{figure:i75} (for the sake of convenience, 
the separate direct and returning radiation contributions
are shown in the polarization observable plots only).}
\label{figure:varitau}
\end{center}
\end{figure*}
In order to better understand how returning radiation 
influences the polarization properties of collected 
photons at infinity and to disentangle its effects 
from those of direct radiation, we performed analogous 
simulations assuming that photons are emitted from 
the disk as unpolarized (i.e. we artificially imposed 
$P^\dir_\loc=0$). Illustrative plots are shown in 
Figure \ref{figure:directoff} for $a=0$, $0.9$ and 
$0.998$ and $i=75^\circ$ (red lines), while the 
corresponding ones, i.e. for direct radiation polarized 
according to Chandrasekhar's (\citeyear{chan60}) 
formulae (see Figure \ref{figure:i75}), are reported 
in blue for ease of comparison. As it can be easily 
observed in the polarization fraction plots (middle 
row), returning photons are practically polarized at 
the same degree in the two cases, with only slight 
differences $\la1\%$ (which are larger for smaller 
BH spins). This shows that the polarization properties 
of returning radiation do not actually depend on the 
intrinsic polarization state of photons. Rather, 
polarization of returning photons rises essentially 
upon reflection at the disk surface (see Appendix 
\ref{appendix:drformulae}). Moreover, assuming that 
photons are unpolarized at the emission, the collected 
radiation turns out to be polarized along the disk 
axis over the entire energy range, with no transitions 
(red solid lines in the bottom row of Figure \ref{figure:directoff}). 
This allows the total polarization degree to monotonically
increase with the photon energy (red solid lines in
the middle row), up to reach a modest degree of polarization 
at sufficiently high energies which can also exceed 
that expected when direct photon polarization is properly 
accounted for.

\subsection{Exploring different optical depths} \label{subsection:differenttau}
\label{section:differenttau}
While the results discussed in the previous subsection 
have been obtained assuming an infinite optical depth 
$\tau$ for the scattering atmosphere on the top of the
blackbody-emitting disk, here we investigate how the
spectral and polarization properties of the collected
radiation can be influenced by changing $\tau$. We remind,
in this regard, that special tables generated through 
the Monte Carlo code {\sc stokes} are used to evaluate 
the photon Stokes parameters when finite values of $\tau$ 
are considered, at variance with the case of infinite 
$\tau$, for which the photon polarization state can be 
obtained using the analytical formulae by \citet{chan60}.
Figure \ref{figure:varitau} shows the 
behaviors of spectra and polarization observables at 
infinity, plotted as functions of the photon energy, 
for a maximally-rotating BH ($a=0.998$)\footnote{Results 
for different values of the BH spin are qualitatively 
similar and are not shown.} and different inclinations 
of the observer's line-of-sight. Results for $\tau=5$, 
$2$, $1$, $0.5$ and $0.2$ are provided, together with 
the case of $\tau\rightarrow\infty$ (black lines),
already discussed in \S \ref{subsection:100albedo}.

\begin{figure*}
\begin{center}
\includegraphics[width=17.5cm]{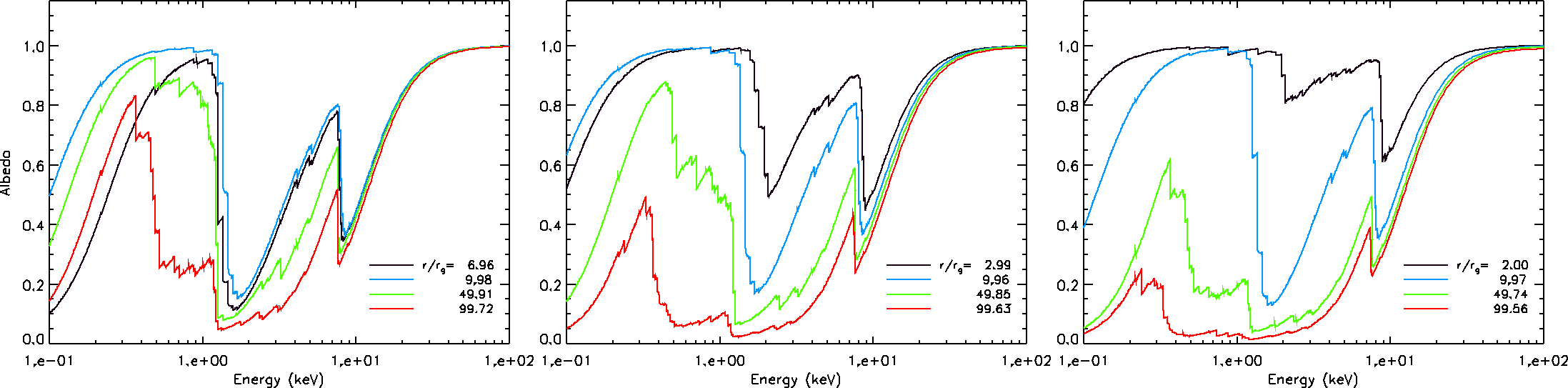}
\caption{Albedo profile as a function of the local photon 
energy $E_\loc$ obtained from {\sc cloudy} according to the 
temperature and density profiles given by equations (\ref{equation:ntprofile}) 
and (\ref{equation:verticaldensity}), respectively, with
$N_{\rm H}=10^{24}$ cm$^{-2}$, for $a=0$ (left), $0.9$ 
(center) and $0.998$ (right). Different colors refer to
different radial distances from the center (see the plot
legend for the color code). Values of mass and accretion
rates are taken as in Figure \ref{figure:i75}.}
\label{figure:albedoprofilesvaria}
\end{center}
\end{figure*}
\begin{figure*}
\begin{center}
\includegraphics[width=17.5cm]{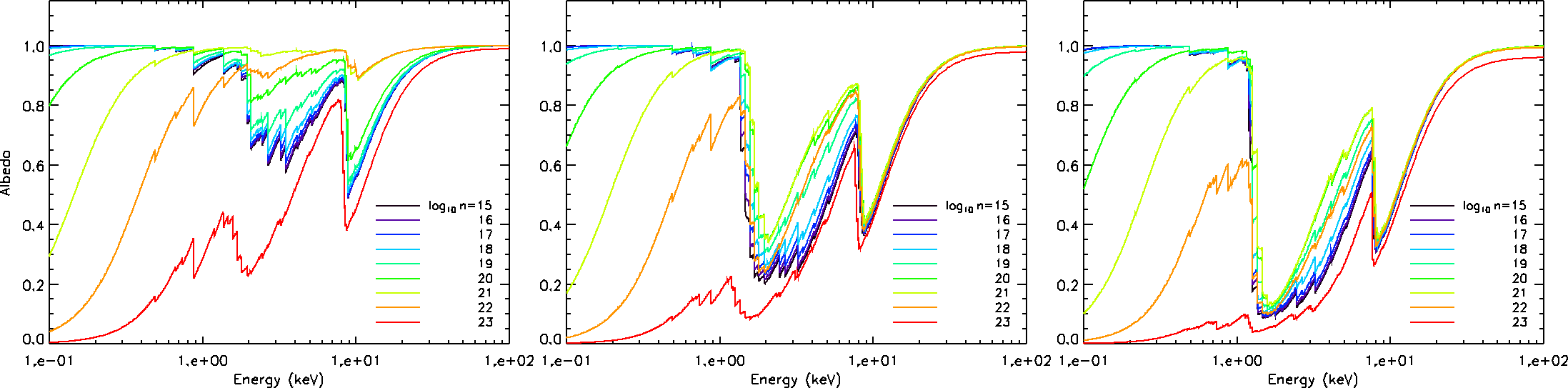}
\caption{Albedo profile as a function of the local photon 
energy $E_\loc$ obtained from different runs of {\sc cloudy} 
according to the temperature profile given by equation (\ref{equation:ntprofile}),
for $a=0.998$ and $r/r_{\rm g}\sim 2$ (left), $5$ (center)
and $10$ (right). The total hydrogen density is considered
as constant for each run, with $\log n(\rm H)=15$--$23$,
step $1$, in units of cm$^{-3}$ (see the plot legend for
the color code). $N_{\rm H}=10^{24}$ cm$^{-2}$, while the 
values of mass and accretion rates are as in Figure \ref{figure:i75}.}
\label{figure:albedoprofilesvariedens}
\end{center}
\end{figure*}
\begin{figure*}
\begin{center}
\includegraphics[width=17.5cm]{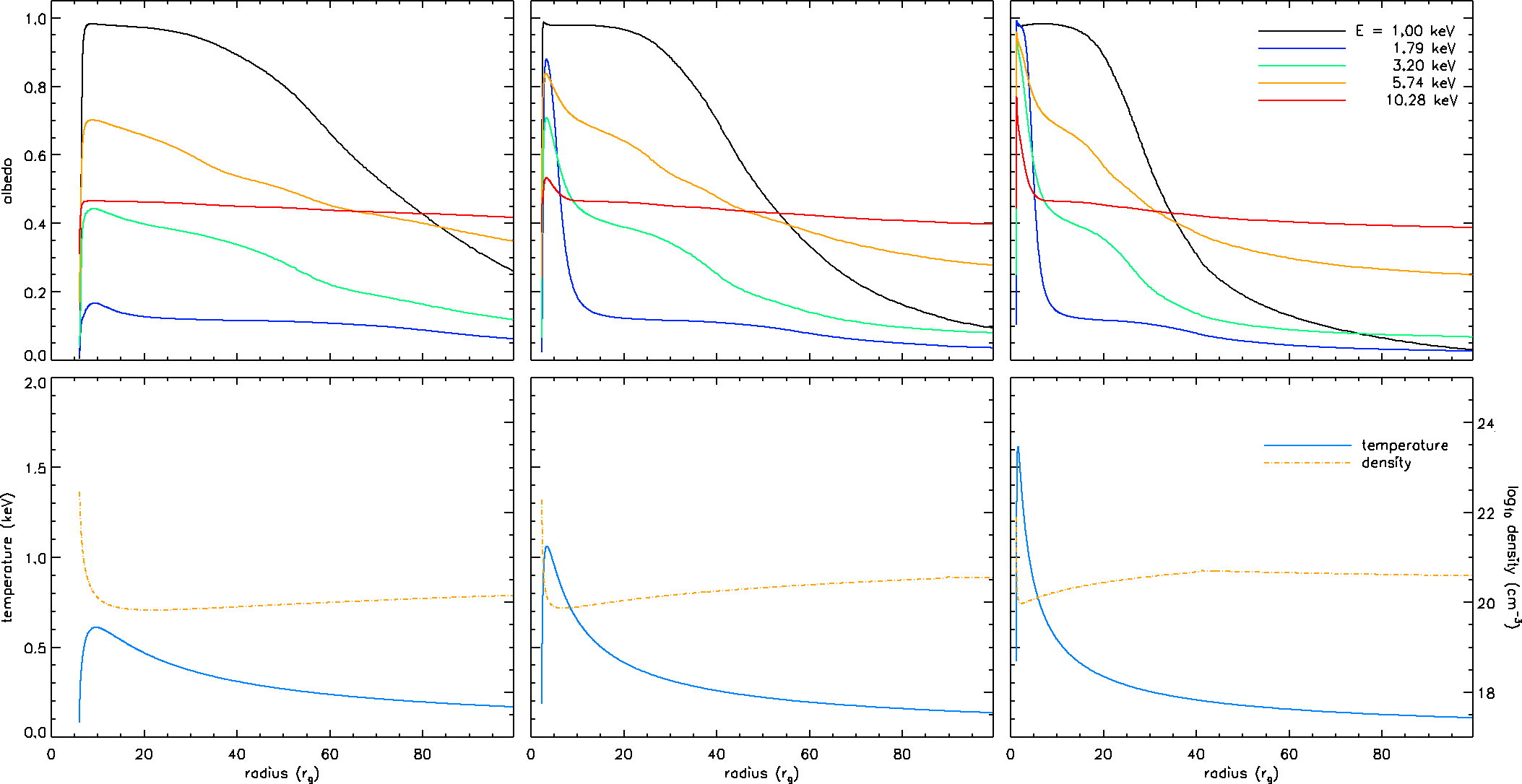}
\caption{Top row: albedo profile as a function of the 
radial distance $r$, obtained for different local energies 
in the $1$--$10$ keV range (see the plot legend for
the color code), in the cases of $a=0$ (left), $0.9$ 
(center) and $0.998$ (right). Bottom row: temperature 
(blue, solid line) and density (orange, dashed line) 
radial profiles described by equations (\ref{equation:ntprofile}) 
and (\ref{equation:verticaldensity}), respectively, 
for $a=0$ and $\dot{M}=24.5\times10^{17}$ g s$^{-1}$ 
(left), $a=0.9$ and $\dot{M}=9\times10^{17}$ g s$^{-1}$ 
(middle), $a=0.998$ and $\dot{M}=3.5\times10^{17}$ 
g s$^{-1}$ (right). In all the plots a BH mass of 
$M=10\,M_\odot$ has been assumed.}
\label{figure:albedovsEvsR}
\end{center}
\end{figure*}
Spectra (top row) turn out to be quite unaffected by 
changing the optical depth; an appreciable, albeit 
small, spread is visible only in the case of $i=75^\circ$,
mainly due to the different emission directionality 
of the primary emission. However, the most considerable 
changes appear in the polarization observables of 
the direct radiation components (see dotted lines 
in the middle and bottom rows). If on one hand both 
the polarization degree and angle behaviors for $\tau=5$ 
(purple) resemble those obtained for an infinite optical 
depth (black), on the other hand quite different 
trends are displayed for the other cases referred. 
More in detail, the polarization angle (bottom row) 
turns out to assume a value close to $0$ over the 
entire energy range for $\tau\la2$. This can be explained 
by the fact, already noticed by \citet[see 
their Figure 1]{dov+08}, that a transition 
occurs at $\tau\sim2$ in the orientation of 
the polarization vectors for direct photons: 
polarization is perpendicular to the projected 
disk symmetry axis ($\chi^\dir_\obs=90^\circ$) 
at high values of $\tau$, while it becomes parallel 
($\chi^\dir_\obs=0^\circ$, as for returning radiation) 
at lower optical depths. An overview of the three 
plots, reported in the bottom row of Figure 
\ref{figure:varitau}, shows that this transition 
also depends on the inclination angle. In fact, 
focussing on the case of $\tau=2$ (cyan), 
$\chi^\dir_\obs\sim 0^\circ$ for $i=45^\circ$ 
and $60^\circ$ at low energies (it decreases 
towards negative values at higher energies due 
to general relativistic effects, see \S 
\ref{subsection:100albedo}), while it becomes 
significantly larger than $0$ only for $i=75^
\circ$. By contrast, the polarization angle 
of returning radiation only (dashed lines) 
remains basically unchanged with respect to 
the case of $\tau\rightarrow\infty$ also when 
a finite optical depth is considered. This
naturally follows from the fact that the
polarization properties of returning photons 
barely depend on the mechanisms that polarize
photons at their emission, being essentially 
determined by reflection (as discussed above). 
As a result, the polarization angle related 
to the joint contribution (direct $+$ returning 
radiation, solid lines) shows the typical swing 
due to the transition between the two regimes 
(see Figures \ref{figure:i75}--\ref{figure:i45}) 
only for $\tau=5$ and $\tau\rightarrow\infty$, 
while no transition occurs for lower optical 
depths.

Looking at the polarization fraction plots (middle 
row) one can observe a behavior with optical depth 
and inclination angle similar to that just discussed 
for the polarization angle. In particular, the 
polarization degree for direct radiation ($P^\dir
_\obs$) at $\tau=2$ and $i=75^\circ$ attains very 
low values ($<0.1\%$) at low energies, before to 
increase at higher energies up to a value not much 
larger than $1\%$. On the contrary, for lower 
inclination angles ($i=45^\circ$ and $60^\circ$) 
$P^\dir_\obs$ maintains a value of the order of 
few precents in the entire energy range. For even 
smaller values of the optical depth, the polarization 
fraction of the direct component attains generally 
higher values, regardless of the observer's inclination. 
This is mainly due to the increase in polarization 
fraction that direct photons undergoes by decreasing 
$\tau$, as again confirmed by the results discussed 
in \citet{dov+08}. On the other hand, much in the same 
way as the polarization angle, the polarization degree 
of returning radiation is quite independent on the 
variation of optical depth, apart from a discrete 
increase at low energies ($\sim 2$--$5\%$) by 
decreasing $\tau$, due essentially to the higher
degree of polarization which characterize photons
at their emission.

Summing together the contributions of direct and returning 
radiation, the total polarization fraction turns out to 
increase in general by decreasing the optical depth, due 
essentially to the growth in $P^\dir_\obs$ at lower energies. 
The only exception occurs considering $\tau=2$. In this case 
the almost unpolarized direct component forces $P^\tot_\obs$ 
to assume a value even smaller than for $\tau\ga5$ at lower
energies, following a trend similar to that followed by
the red solid line in the middle-right panel of Figure 
\ref{figure:directoff} (where indeed $P^\dir_\obs$ is 
forced to be zero). 

\subsection{More realistic albedo prescription}
\label{subsection:albedo}
\begin{figure*}
\begin{center}
\includegraphics[width=17.5cm]{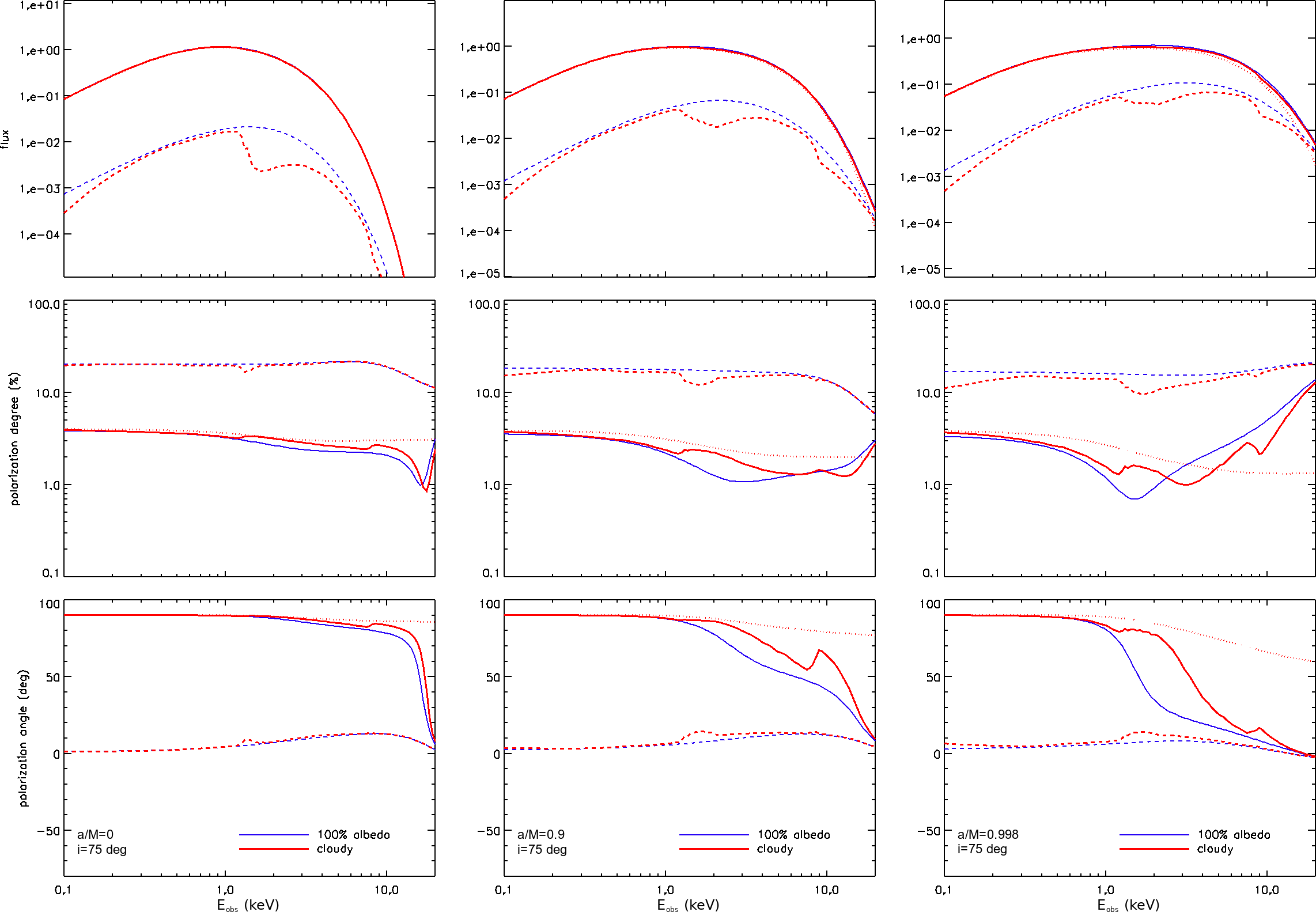}
\caption{Spectrum (top row), polarization degree (middle
row) and polarization angle (bottom row) behaviors, plotted 
as functions of the photon energy at the observer, for 
a $100\%$ albedo prescription (blue lines) and assuming 
the albedo profile obtained from {\sc cloudy} (red lines)
using the \citet{nt73} temperature profile (equation 
\ref{equation:ntprofile}), the \citet{co17} density
profile (see Appendix \ref{appendix:density}) and 
for $N_{\rm H}=10^{24}$ cm$^{-2}$. The values of 
the other parameters and line styles are taken as 
in Figure \ref{figure:i75}.}
\label{figure:albedoi75}
\end{center}
\end{figure*}
\begin{figure*}
\begin{center}
\includegraphics[width=17.5cm]{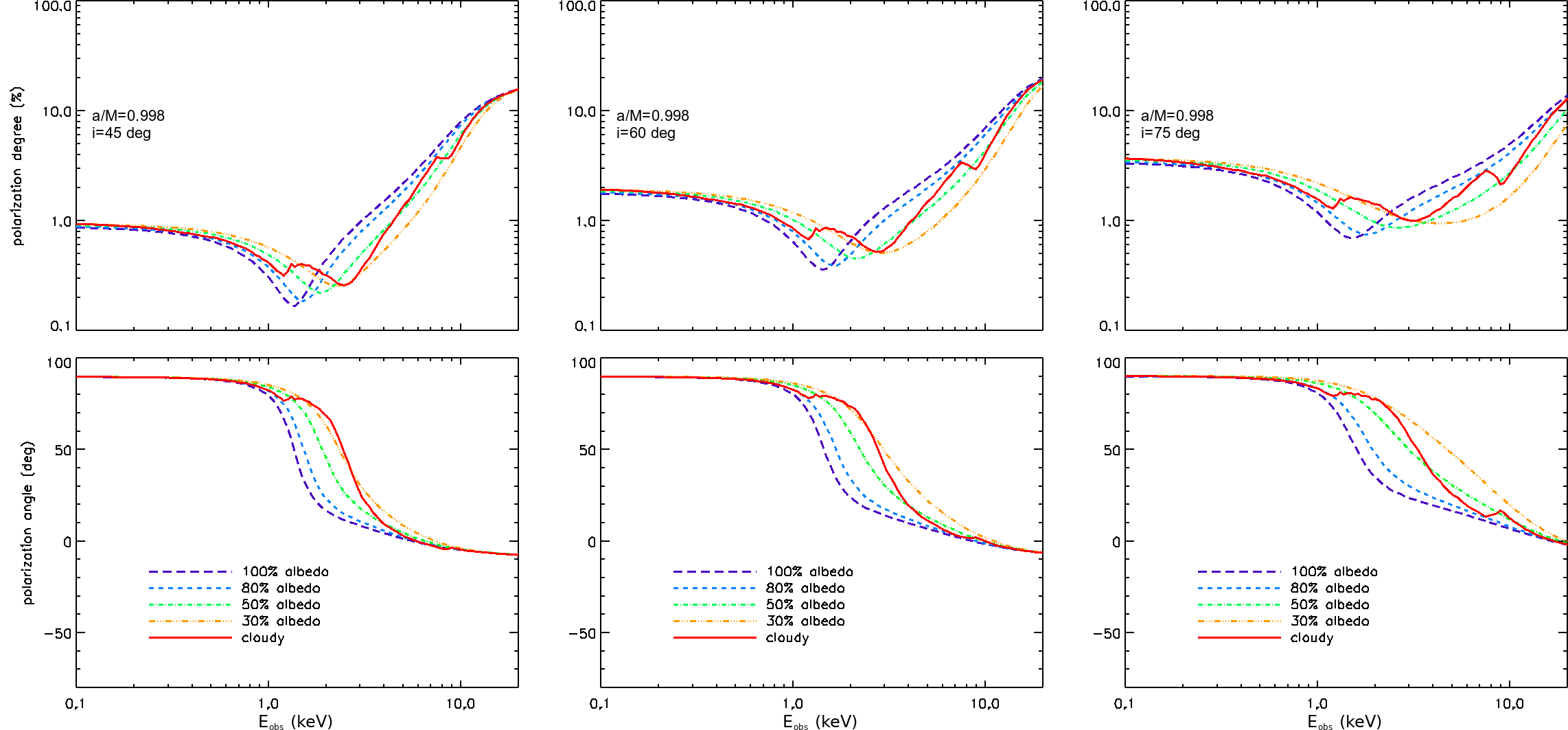}
\caption{Joint (direct $+$ returning radiation) polarization 
degree (top row) and angle (bottom row) plotted as functions 
of the photon energy at the observer for constant $100\%$ 
(blue, long dashed), $80\%$ (cyan, short dashed), $50\%$ 
(green, dash-dotted) and $30\%$ (orange, dash-double dotted)
albedo prescriptions, as well as for the albedo profile 
produced by {\sc cloudy} (red, solid), reported in Figures 
\ref{figure:albedoprofilesvaria}--\ref{figure:albedovsEvsR}. 
The BH spin is taken as $0.998$, while the observer's line-of-sight 
inclination is $45^\circ$ (left), $60^\circ$ (center) and $75^
\circ$ (right). The values of the other parameters are 
taken as in Figure \ref{figure:i75}.}
\label{figure:varialbedivarii}
\end{center}
\end{figure*}
A critical simplification we assumed so far in 
our work is that of considering a constant, $100
\%$ albedo prescription at the disk surface (which 
implies that returning photons are all reflected 
towards the observer). Although computing a self-consistent 
ionization profile for the disk is beyond the scope 
of this paper (and it will be addressed in a future 
work), here we tried to relax the constraint on 
the albedo, at least in a simplified way, with 
the purpose to convey a sense about the effects 
that can be produced on spectra and polarization 
observables when these aspects are properly taken 
into account.

\subsubsection{Albedo profile} \label{subsubsec:albedoprofile}
In order to calculate a more realistic albedo profile for 
the disk surface, we use the version 17.00 of {\sc cloudy}, 
last described by \citet{cloudy}, a code to simulate the 
micro-physical processes that occur in astrophysical clouds, 
allowing for the prediction of the spectral properties of 
the radiation field that emerges (or is reflected from) these 
clouds. More in detail, we exploit the {\tt coronal} setup, 
in which the gas which composes the medium is assumed to be 
ionized due to mutual collisions. Input parameters of each 
run are the (constant) gas kynetic temperature $T$ and the 
total (i.e. ionic, atomic and molecular) hydrogen density 
$n({\rm H})$. One should also provide to the code a value 
of the hydrogen column density $N_{\rm H}$ at which calculations 
are stopped. In this respect, we take in the following $N_{\rm 
H}=10^{24}$ cm$^{-2}$, which corresponds to $\tau\sim 1$ 
for Compton scattering (i.e. the process we are mainly interested 
for in the disk); this is tantamount to consider the layer 
in which most of scatterings happen. The output is a text 
file containing the values of the scattering, absorption and 
total opacities and the albedo parameter as functions of 
the energy, for the values of temperature and densities 
specified in input. For coherence with the results discussed 
above, we use the \citet{nt73} profile given in equation 
(\ref{equation:ntprofile}) to describe the variation of 
the temperature with the radial distance from the center. 
For the density, instead, we used the formulae reported 
in \citealt{co17}, which give the local density profile 
in the inner, middle and outer regions of a general 
relativistic, standard disk (see Figure \ref{figure:albedovsEvsR}, 
bottom row, and Appendix \ref{appendix:density} for more 
details).

The albedo profile is then computed by {\sc cloudy} after 
having specified the values of temperature $T(r)$ and density 
$\rho(r)$ for each radial patch (with $r$ the radial distance). 
Results for the energy-dependent albedo profile $A(E_\loc)$
obtained for different radial patches in the cases of $a=0$
(left), $0.9$ (center) and $0.998$ (right) are shown in 
Figure \ref{figure:albedoprofilesvaria}. As it can be clearly 
seen, in all the cases explored the simplifying assumption 
of 100\% albedo turns out to be a good approximation only 
at very high energies, i.e. $E_{\rm loc}\approx 10$--$100$ 
keV. Elsewhere, the albedo significantly deviates from 
unity (except for some values of $r$ at lower energies),
especially in the $0.1$--$10$ keV band, which is indeed 
the working energy range of the forthcoming X-ray polarimeters 
like {\it IXPE}. Here different line features appear, 
more or less visible depending on the BH spin and radial 
distance, such as the clear iron absorption edge which 
occurs at around $\sim 6$--$7$ keV \cite[see e.g.][]{yrf98}.
Plots in Figure \ref{figure:albedoprofilesvariedens} give 
a more exhaustive view on how the albedo profile depends 
on the density of the slab in which calculations are performed. 
In this case the outputs are obtained for $a=0.998$, three 
different values of the temperature, i.e. those corresponding 
to $r=2$ (left), $5$ (middle) and $10\,r_{\rm g}$ (right)
according to the \citet{nt73} temperature profile, and different 
values of the total hydrogen density $n({\rm H})$ between 
$10^{15}$ and $10^{23}$ cm$^{-3}$. The plots show that 
the dependence of $A(E_\loc)$ on the density is stronger 
at low energies, where it exhibits an increasing behavior 
by decreasing $n({\rm H})$. In particular, it attains
values close to $0$ at around $0.1$ keV for particle 
densities in excess of $10^{22}$--$10^{23}$ cm$^{-3}$, 
especially for lower temperatures (i.e. for larger 
radial distances, see the right panel). On the other 
hand, at higher energies ($\ga10$--$20$ keV) the albedo 
tends to reach the same value ($\approx 1$) in the 
entire range of densities explored.

The top row of Figure \ref{figure:albedovsEvsR} 
finally shows the {\sc cloudy} albedo profile as a 
function of the radial distance $A(r)$ for the same 
three values of $a$ discussed in Figure \ref{figure:albedoprofilesvaria}
and assuming the temperature and density profiles
expressed by equations (\ref{equation:ntprofile}) 
and (\ref{equation:verticaldensity}), resepctively.
The outputs for five different local energies, covering
the $1$--$10$ keV band, are shown. Also in this case
an overall reduction of the albedo with respect to $100\%$, 
is shown, with in general a decreasing behavior as a 
function of the radial distance (with the only exception 
at $E_\loc\approx 10$ keV, where $A(r)$ is essentially
constant at $\approx 0.4$ over the entire radial range 
considered. A maximum, where the albedo attains a value 
close to $1$, can be observed at energies $E_\loc\sim 1$
keV as the innermost stable circualr orbit is approached, 
essentially in correspondence with the peaks of both the 
temperature and density profiles (see bottom row). This 
maximum looks rather broadened for energies close to $E_
\loc\approx1$ keV, where the albedo remains at around 
$100\%$ up to a distance of $20$--$30\,r_{\rm g}$ from 
the center.

\subsubsection{Results} \label{subsubsec:includingalbedo}
For coherence with the results discussed previously 
(see \S \ref{subsection:100albedo} and \ref{subsection:differenttau}),
the albedo profiles obtained from {\sc cloudy} are 
firstly interpolated over the energy grid used 
for {\sc kyn} simulations and then stored in different 
fits files, according to the value of the BH spin. 
Eventually, returning radiation Stokes parameters 
are convolved with the correspondent value of the 
albedo for each energy and radial bin, so that, at 
each local energy $E_\loc$, it is
\begin{flalign} \label{equation:albedoconvolution}
i_\loc(r,\phi)&=i^\dir_\loc(r,\phi)+i^\ret_\loc(r)A(r) \nonumber & \\
q_\loc(r,\phi)&=q^\dir_\loc(r,\phi)+q^\ret_\loc(r)A(r) \nonumber & \\
u_\loc(r,\phi)&=u^\dir_\loc(r,\phi)+u^\ret_\loc(r)A(r)\,, &
\end{flalign}
where $i_\loc$, $q_\loc$ and $u_\loc$ are the energy-dependent
Stokes parameters defined in \S \ref{subsec:returning}.

To provide an example of how including a more realistic 
albedo profile can modify the behaviors of spectral and 
polarization properties of radiation collected from stellar-mass 
BH accretion disks, we report in Figure \ref{figure:albedoi75} 
(red lines) the results obtained for the same values of 
the input parameters as in Figure \ref{figure:i75}, where 
the albedo was assumed to be $100\%$ for every energy 
and radial bin (behaviors of Figure \ref{figure:i75} 
are also reported and marked in blue in Figure 
\ref{figure:albedoi75} for ease of comparison). 
It can be noticed that, although returning photon 
spectra turn out to be strongly affected by the inclusion 
of a more complex albedo profile, total spectra are 
practically unchanged. This follows from the fact that 
the ratio between the returning and the direct photon
numbers is quite small ($\approx10^{-2}$--$10^{-3}$) 
over the almost entire energy range considered. As
noticed in \S \ref{subsection:100albedo}, the only 
exception occurs at very high energies ($\ga10$ keV), 
where photons are mostly emitted from regions closer 
to $\rms$ (see \S \ref{subsection:100albedo}). However, 
as shown in Figure \ref{figure:albedovsEvsR}, for 
$r\rightarrow\rms$ and $E_\loc\ga 10$ keV the albedo 
attains values close to $100\%$ at essentially any 
of the BH spins considered. For this reason, small 
effects on the spectra are anyway expected also at 
high values of $E_\obs$. Effects on polarization 
degree and angle trends are instead more pronounced. 
This is especially visible for higher BH spins ($a=0.9$ 
and $a=0.998$), when returning radiation contributions 
start to be important already at $E_\obs\sim1$ keV. 
More in details, looking at the polarization angle 
plots (bottom row), the most important difference 
with respect to the $100\%$ albedo case concerns 
the swing which occurs when returning radiation 
contributions start to be dominant over the direct 
radiation ones. This appears to be rather broadened 
and in general shifted towards higher energies when
the {\sc cloudy} albedo profile is included. This 
also impacts the behavior of the polarization degree 
(middle row), with the minimum which is moved as well 
to higher energies.

In order to better investigate how indeed these particular
features (both discussed in \S \ref{subsection:100albedo}) 
in the polarization observable plots are modified by 
considering a more realistic albedo prescription, we 
report in Figure \ref{figure:varialbedivarii} the energy-dependent 
behaviors of polarization degree and angle as observed
at infinity for different albedo profiles: constant $100\%$, 
$80\%$, $50\%$ and $30\%$ together with that obtained 
in output from cloudy (see Figure \ref{figure:albedoprofilesvaria}). 
Here the only case of $a=0.998$ is explored, with different 
observer's inclinations ($i=45^\circ$, $60^\circ$ and 
$75^\circ$).While no significant changes occur by changing 
$i$, it clearly turns out that the effect on the polarization 
observables of reducing the albedo at the disk surface 
is mainly that of moving the minimum of the polarization 
degree and the swing of the polarization angle towards 
higher energies. This can be easily explained by the 
fact that convolving an albedo smaller than $1$ to the 
returning photon Stokes parameters acts in further suppressing 
the contribution of returning radiation with respect to 
that of the direct one (see e.g. the top row of Figure 
\ref{figure:albedoi75}). Hence, the energy at which returning 
radiation starts to dominate over the direct component 
consequently raises, moving forward the minimum of the 
polarization degree and the transition in the polarization 
angle.

\begin{figure*}
\begin{center}
\includegraphics[width=17.5cm]{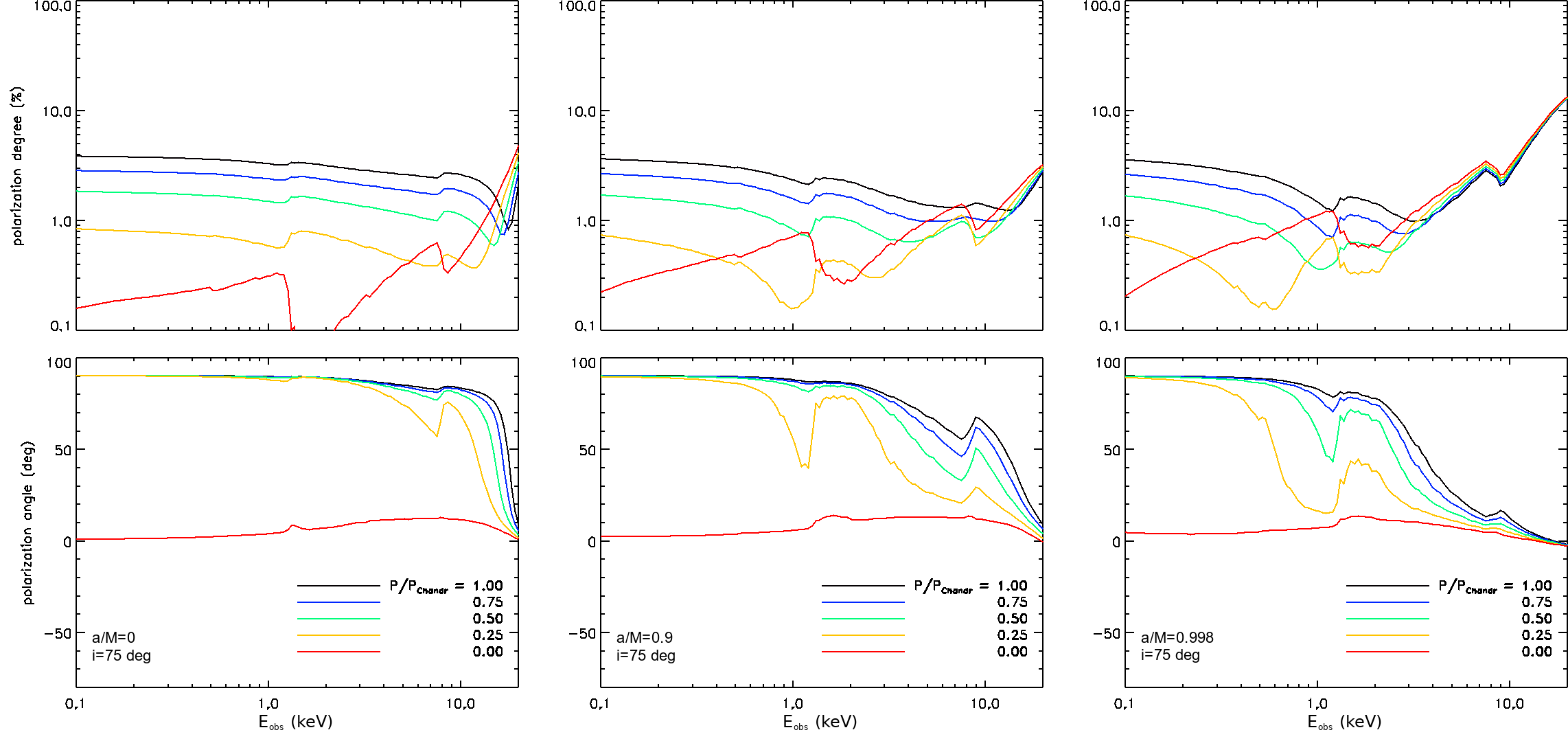}
\caption{Joint (direct $+$ returning radiation) polarization 
degree (top row) and angle (bottom row), plotted as functions 
of the photon energy at the observer, for the same albedo 
profile used in Figure \ref{figure:albedoi75} and for direct 
radiation polarized at different degrees: $P^\dir_\loc=P_{\rm 
Chandr}$ (black), $0.75P_{\rm Chandr}$ (blue), $0.5P_{\rm Chandr}$ 
(green), $0.25P_{\rm Chandr}$ (orange) and $0$ (red), where 
$P_{\rm Chandr}$ denotes the polarization pattern expected 
from \citet{chan60}. The values of the input parameters are 
taken as in Figure \ref{figure:i75}.}
\label{figure:albedoplotdiffP}
\end{center}
\end{figure*}
As well as affecting returning radiation, the absorption 
caused by the ionization of the disk material can be 
reasonably expected to change also the polarization
properties of the emerging radiation. Remarking once
more that a complete treatment of ionization in the 
disk is outside the scope of this paper, we resort to 
artificially reduce the polarization degree of direct 
photons in order to mimic this possible effect. To this
aim we performed a number of simulations for which the
direct radiation polarization degree ranges from $P^\dir
_\loc=P_{\rm Chandr}$ (i.e. following Chandrasekhar's,
\citeyear{chan60}, formulae) down to $0$ (i.e. unpolarized 
emerging light). Results, obtained for the same values 
of parameters assumed in Figure \ref{figure:albedoi75}, 
are illustrated in Figure \ref{figure:albedoplotdiffP}, 
which shows the energy-dependent polarization observables 
as observed at infinity. As expected, the observed polarization 
degree turns out to decrease, in general, by decreasing 
the polarization fraction of the emerging radiation. 
However, when photons emitted from the disk are originally 
unpolarized, the collected radiation may counter-intuitively 
result even more polarized than for non-zero intrinsic 
polarization. This behavior, already discussed in \S
\ref{subsection:100albedo} commenting Figure \ref{figure:directoff}, 
can be clearly seen looking at the red curves in the 
top row of Figure \ref{figure:albedoplotdiffP}. An 
explanation can actually be extracted from the polarization 
angle plots (bottom row), where, as noticed for those 
reported in Figures \ref{figure:albedoi75} and 
\ref{figure:varialbedivarii}, the addition of a non-trivial 
albedo profile broadens the range in energy over which 
the polarization angle swing extends. Since, on the 
other hand, the polarization angle is rather constant 
when direct radiation is assumed to be unpolarized (so 
that no transition occurs at all), the correspondent 
polarization fraction is much less reduced than for 
the other cases, in which instead the variation of 
the polarization angle with the energy is more significant.

\section{Discussion and conclusions}
\label{section:discussion}
In this work we have revisited the problem of the spectral 
and polarization properties of radiation emitted from 
stellar-mass BH accretion disks in the soft state, considering 
the contribution of returning radiation (i.e. photons 
which are bent by the strong BH gravity to return 
to the disk before being reflected towards the observer) 
alongside that of direct radiation (i.e. photons that 
arrive directly to the observer once emitted from the 
disk surface). To this aim we first added to the {\sc 
kyn} package \cite[][]{dov04} a specific module, exploiting
the {\sc c++} code {\sc selfirr} \cite[based on the ray-tracing
{\sc sim5} package, see][]{bur17} to calculate all the possible
null geodesics along which photons can travel between two 
different points on the disk surface (assuming a Kerr space-time, 
see section \ref{section:themodel}). 
Secondly, we have checked that the results of our simulations 
were compatible with those discussed in previous works 
\cite[e.g.][]{sk09,kraw12} which already considered 
returning radiation in their calculations. As widely
discussed in section \ref{subsection:100albedo}, the 
comparison has revealed an overall good agreement over 
a large range of input parameters, with only some discrepancies
in the case of $a=0.9$, occurring essentially at an energy 
higher then $20$ keV. We remarked that statistics at 
such high energies would be definitively too low
for the present polarimetry techniques \cite[see e.g.][]{cos+01,
bell+10} to provide conclusive results.

Using the Monte Carlo code {\sc stokes}, we have been able
to investigate how spectral and polarization properties are 
modified by varying the optical depth of the (pure-scattering) 
atmosphere assumed to cover the disk (see \S \ref{subsection:differenttau}). 
We found that the trends of the polarization observables 
are rather similar to those obtained for an infinite optical 
depth when $\tau\ga 5$. On the other hand, a significantly 
higher polarization fraction (even by a factor of $\sim 6$--$10$ 
at low energies) can be expected when values of $\tau$ smaller 
than $1$ are considered; a transition between the two regimes 
occurs at around $\tau\sim2$ \cite[in agreement with the results 
already discussed by][]{dov+08}. This certainly changes the 
expected polarization signature of soft-state BH accretion 
disks in the $2$--$10$ keV energy range, which will be attained 
by the next-generation polarimeters like {\it IXPE} \cite[][]{weiss+13}. 
Moreover, if future polarimeters capable to investigate 
polarization also at energies $\approx 0.1$--$2$ keV \cite[like 
{\it XPP}, see][]{kraw+19} will be deployed in the next 
years, this behavior can be used in principle to obtain 
information on the optical properties of the medium from 
which photons have been emitted.

Without entering into details (a more complete treatment 
of the ionization profile in the disk will be addressed 
in a future work), we then started to make a more realistic 
assumption for the disk albedo profile, going beyond the 
simplified prescription of taking it constant at $100\%$. 
To this aim we used the code {\sc cloudy} \cite[][]{cloudy}, 
assuming that ionization in the disk material is due essentially 
to mutual collisions between its particles. The albedo
profiles obtained from {\sc cloudy} were then convolved 
with the returning radiation Stokes parameters computed 
by {\sc kyn}. The plots reported in \S \ref{subsubsec:includingalbedo}
offer an illustration of the effects of a more realistic 
albedo prescription on both spectra and polarization observables.
A comparison with the case discussed in section \ref{subsection:100albedo} 
shows that introducing in the calculations a more complicated 
albedo profile can significantly alter the behavior of polarization 
degree and angle with respect to the simplifying $100\%$-albedo 
prescription. This can in particular affect the constraints 
on BH spin and inclination angle which may be extracted from 
polarization measurements (see Figure \ref{figure:varialbedivarii}).

Finally, we tried to reproduce the possible effects 
of the disk ionization also on the polarization properties 
of the emerging radiation. As a first attempt (which 
will be also investigated more in depth in a future 
work), we resorted to reduce the polarization degree 
assigned to the direct radiation component in our simulations, 
starting from the pure-scattering polarization pattern 
predicted by \citet{chan60} up to assuming an unpolarized 
radiation (see Figure \ref{figure:albedoplotdiffP}). 
Such effects turned out to be indeed quite important,
especially in the case of weakly-polarized direct 
photons, for which radiation collected at infinity 
may result even more polarized with respect to the 
case of Chandrasekhar-like intrinsic polarization.

To conclude, on the wave of previous works \cite[see
e.g.][]{dov+08,sk09,kraw12,sk13}, we confirm that
X-ray polarization measurements can be crucial in
extending our knowledge about stellar-mass BH accretion
disks in the soft state. However, when the optical 
structure of the disk is considered in more detail
(e.g. taking the surface optical depth and the disk 
ionization as free parameters of the problem), the 
simple scenario depicted under the assumptions of 
$\tau=1$ and $100\%$ albedo may be not the right 
description. However, we point out that there
are several effects, we did not consider in the 
present work, which can somehow mitigate the degeneracy
introduced by considering absorption in the disk
material. For example, the contributions due to
magnetic pressure in determining the vertical structure
of the disk or the photoionization due to radiation
coming from an external corona, may significantly
increase the ionization fraction and therefore
increase the albedo as well. Moreover, deeper 
investigations, which are outside the scope of 
the present paper, are requested in order to provide 
a more realistic picture of these kind of sources 
through the joint effort of spectroscopy and X-ray 
polarimetry. This involves a self-consistent treatment 
of ionization in the disk material, a more general 
prescription for the vertical structure of the disk 
(see Appendix \ref{appendix:density}) and the inclusion 
of absorption effects in the radiative transfer 
calculations of polarized thermal photons (i.e. 
not only for the reflected ones). We are planning 
to address these issues in future works.

The sensitive functional dependence of the
polarization observables (seen as functions of the
photon energy) on the physical state of the accretion 
disk should allow us to set more stringent constraints 
on the models of the accretion medium along with 
the effects of General Relativity. This makes it 
possible in future, also exploiting the capabilities
of instruments like {\it IXPE}, to reduce the inherent 
degeneracies of these models and to measure the parameters, 
including the BH spin.

\section*{Acknowledgments}
We thank the anonymous referee for his/her constructive
comments which helped in improving a previous version of
this paper. RT thanks Roberto Turolla for some helpful 
discussions. RT, SB and GM acknowledge financial support 
from the Italian Space Agency (grant 2017-12-H.0). MD, 
MB and VK aknowledge the support by the project RVO:67985815 
and the project LTC18058. WZ would like to thank GACR 
for the support from the project 18-00533S.


\appendix
\section{Diffuse reflection formulae} \label{appendix:drformulae}
Chandrasekhar's (\citeyear{chan60}) diffuse reflection law can be 
expressed as
\begin{flalign} \label{equation:diffreflgen}
\left(\begin{array}{c}
I_{\rm l} \\ I_{\rm r} \\ U
\end{array}\right)&=\frac{1}{4\mu}\boldsymbol{Q}
\boldsymbol{S}(\mu,\varphi;\muz,\varphiz)\left(\begin{array}{c}
F_{\rm l} \\ F_{\rm r} \\ F_{\rm U}
\end{array}\right)\,, &
\end{flalign}
where $I_{\rm l}$, $I_{\rm r}$ and $U$ are the (number) 
intensities which characterize the radiation field ($l$ 
and $r$ refer to two mutually orthogonal directions, in
the plane made by the disk symmetry axis and the observer's
line-of-sight and perpendicular to this plane, respectively), 
$F_{\rm l}$, $F_{\rm r}$ and $F_{\rm U}$ are the correspondent 
(number) fluxes, $\mu=\cos\theta$ ($\muz=\cos\bar{\theta}_{\rm 
i}$) is the cosine of the emission (incidence) angle 
that the propagation direction makes with the disk 
symmetry axis and
\begin{flalign} \label{equation:Qmatrix}
\boldsymbol{Q}&=\left(\begin{array}{ccc}
1 & 0 & 0 \\
0 & 1 & 0 \\
0 & 0 & 2
\end{array}\right)\,. &
\end{flalign}
The elements of the matrix
\begin{flalign} \label{equation:Smatrix}
\boldsymbol{S}(\mu,\varphi;\muz,\varphiz)&=\frac{3}{4}
\frac{\muz\mu}{\mu+\muz}\left(\begin{array}{ccc}
S_{11} & S_{12} & S_{13} \\
S_{21} & S_{22} & S_{23} \\
S_{31} & S_{32} & S_{33}
\end{array}\right) &
\end{flalign}
are given by the following expressions:
\begin{flalign} \label{equation:S11}
S_{11}&=\psi(\mu)\psi(\muz)+2\phi(\mu)\phi(\muz) \nonumber & \\
\ &-4\mu\muz[(1-\mu^2)(1-\muz^2)]^{1/2}H^{(1)}(\mu)H^{(1)}(\muz)
\cos(\varphiz-\varphi) \nonumber & \\
\ &+\mu^2\muz^2H^{(2)}(\mu)H^{(2)}(\muz)\cos[2(\varphiz-\varphi)]\,; &
\end{flalign}
\begin{flalign} \label{equation:S12}
S_{12}&=\psi(\mu)\chi(\muz)+2\phi(\mu)\zeta(\muz) \nonumber & \\
\ &-\mu^2H^{(2)}(\mu)H^{(2)}(\muz)\cos[2(\varphiz-\varphi)]\,; &
\end{flalign}
\begin{flalign} \label{equation:S13}
S_{13}&=2\mu[(1-\mu^2)(1-\muz^2)]^{1/2}H^{(1)}(\mu)H^{(1)}(\muz)
\sin(\varphiz-\varphi) \nonumber & \\
\ &-\mu^2\muz H^{(2)}(\mu)H^{(2)}(\muz)\sin[2(\varphiz-\varphi)]\,; &
\end{flalign}
\begin{flalign} \label{equation:S21}
S_{21}&=\chi(\mu)\psi(\muz)+2\zeta(\mu)\phi(\muz) \nonumber & \\
\ &-\muz^2H^{(2)}(\mu)H^{(2)}(\muz)\cos[2(\varphiz-\varphi)]\,; &
\end{flalign}
\begin{flalign} \label{equation:S22}
S_{22}&=\chi(\mu)\chi(\muz)+2\zeta(\mu)\zeta(\muz) \nonumber & \\
\ &+H^{(2)}(\mu)H^{(2)}(\muz)\cos[2(\varphiz-\varphi)]\,; &
\end{flalign}
\begin{flalign} \label{equation:S23}
S_{23}&=\muz H^{(2)}(\mu)H^{(2)}(\muz)\sin[2(\varphiz-\varphi)]\,; &
\end{flalign}
\begin{flalign} \label{equation:S31}
S_{31}&=2\muz[(1-\mu^2)(1-\muz^2)]^{1/2}H^{(1)}(\mu)H^{(1)}(\muz)
\sin(\varphiz-\varphi) \nonumber & \\
\ &-\mu\muz^2H^{(2)}(\mu)H^{(2)}(\muz)\sin[2(\varphiz-\varphi)]\,; &
\end{flalign}
\begin{flalign} \label{equation:S32}
S_{32}&=\mu H^{(2)}(\mu)H^{(2)}(\muz)\sin[2(\varphiz-\varphi)]\,; &
\end{flalign}
\begin{flalign} \label{equation:S33}
S_{33}&=[(1-\mu^2)(1-\muz^2)]^{1/2}H^{(1)}(\mu)H^{(1)}(\muz)
\cos(\varphiz-\varphi) \nonumber & \\
\ &-\mu\muz H^{(2)}(\mu)H^{(2)}(\muz)\cos[2(\varphiz-\varphi)]\,. &
\end{flalign}
The values that the special functions $\psi(\mu)$, $\phi
(\mu)$, $\chi(\mu)$, $\zeta(\mu)$, $H^{(1)}(\mu)$ and 
$H^{(2)}(\mu)$ take on a descrete grid of $\mu$ are 
reported in Table XXV of \citet{chan60}. 

Stokes parameters can be obtained from $I_{\rm l}$, $I_{\rm r}$
and $U$ given in equation (\ref{equation:diffreflgen}) through
\begin{flalign} \label{equation:defiuq}
\bar{i}_\refl&=I_{\rm l}+I_{\rm r} \nonumber & \\
\bar{q}_\refl&=I_{\rm l}-I_{\rm r} \nonumber & \\
\bar{u}_\refl&=U\,, &
\end{flalign}
while the components of the Stokes vectors $\boldsymbol{
\bar{s}}_\refl(0,-)$, $\boldsymbol{\bar{s}}_\refl(1,0)$ 
and $\boldsymbol{\bar{s}}_\refl(1,\pi/4)$ used in equation 
(\ref{equation:combinationstokesvector}) can be obtained 
from equation (\ref{equation:diffreflgen}) by placing 
$F_{\rm l}=F_{\rm r}=f_\loc/2$, $F_{\rm U}=0$ for 
unpolarized light, $F_{\rm l}=f_\loc$, $F_{\rm r}=F_{\rm 
U}=0$ for vertically-polarized light and $F_{\rm l}=F_{\rm 
r}=f_\loc/2$, $F_{\rm U}=f_\loc$ for $45^\circ$-polarized 
light.

\section{Disk density profile} \label{appendix:density}
In order to calculate coherently the albedo profile of
the disk with {\sc cloudy}, we adopted the density profile
given by \citet{co17}, who revised the standard disk local 
structure formulae originally discussed by \citet[see also 
\citealt{pt74} and \citealt{af13}]{nt73} for both stellar-mass
and supermassive BHs. For the sake of a better readability,
we reported in the following the main expressions we used
in our calculations. The disk is divided into three main
regions, according to what mechanism dominates the transfer
of radiation in the disk material. In the inner (or edge) 
region, where it holds
\begin{flalign}
\frac{p_{\rm gas}}{p_{\rm rad}}&=(2.6\times10^{-5})\,\alpha^{-1/4}M_*^{7/4}\dot{M}_*^{-2} \nonumber & \\
\ &\ \ \ \ \times x^{29/4}\mathcal{C}^{9/4}\mathcal{D}^{-1/4}\mathcal{R}^{5/4}\mathcal{P}^{-2}\ll 1\,, \nonumber &
\end{flalign}
\begin{flalign} \label{equation:conditioninn}
\frac{\kappa_{\rm ff}}{\kappa_{\rm es}}&=(2.2\times10^{-8})\,\alpha^{-1/8}M_*^{15/8}\dot{M}_*^{-2} \nonumber & \\
\ &\ \ \ \ \times x^{61/8}\mathcal{C}^{17/8}\mathcal{D}^{-1/8}\mathcal{R}^{9/8}\mathcal{P}^{-2}\ll 1\,, &
\end{flalign}
with $p_{\rm gas}$ ($p_{\rm rad}$) the gas (radiation)
pressure and $\kappa_{\rm ff}$ ($\kappa_{\rm es}$) the
free-free (scattering) opacity, the (equatorial) hydrogen 
density $n_0(\rm H)$ is given by
\begin{flalign} \label{equation:ninn}
n_0({\rm H})_{\rm inn}&=(1.50\times10^{19}\,{\rm cm^{-3}})\,\alpha^{-1}M_*\dot{M}_*^{-2} \nonumber & \\
\ &\ \ \ \ \times x^5\mathcal{C}^3\mathcal{D}^{-1}\mathcal{R}^2\mathcal{P}^{-2}\,. &
\end{flalign}
Then, in the middle region, where 
\begin{flalign}
\frac{p_{\rm rad}}{p_{\rm gas}}&=(69)\,\alpha^{1/10}M_*^{-7/10}\dot{M}_*^{4/5} \nonumber & \\
\ &\ \ \ \ \times x^{-29/10}\mathcal{C}^{-9/10}\mathcal{D}^{1/10}\mathcal{R}^{-1/2}\mathcal{P}^{4/5}\ll 1\,, \nonumber &
\end{flalign}
\begin{flalign} \label{equation:conditionmid}
\frac{\kappa_{\rm ff}}{\kappa_{\rm es}}&=(4.4\times10^{-6})\,M_*\dot{M}_*^{-1} \nonumber & \\
\ &\ \ \ \ \times x^{4}\mathcal{C}\mathcal{R}^{1/2}\mathcal{P}^{-1}\ll 1\,, &
\end{flalign}
one has
\begin{flalign} \label{equation:nmid}
n_0({\rm H})_{\rm mid}&=(4.86\times10^{24}\,{\rm cm^{-3}})\,\alpha^{-7/10}M_*^{-11/10}\dot{M}_*^{2/5} \nonumber & \\
\ &\ \ \ \ \times x^{-37/10}\mathcal{C}^{3/10}\mathcal{D}^{-7/10}\mathcal{R}^{1/2}\mathcal{P}^{2/5}\,. &
\end{flalign}
Finally, in the outer region, where
\begin{flalign}
\frac{p_{\rm rad}}{p_{\rm gas}}&=(0.27)\,\alpha^{1/10}M_*^{-1/4}\dot{M}_*^{7/20} \nonumber & \\
\ &\ \ \ \ \times x^{-11/10}\mathcal{C}^{-9/20}\mathcal{D}^{1/10}\mathcal{R}^{-11/40}\mathcal{P}^{7/20}\ll 1\,, \nonumber &
\end{flalign}
\begin{flalign} \label{equation:conditionout}
\frac{\kappa_{\rm es}}{\kappa_{\rm ff}}&=(4.8\times10^{2})\,M_*^{-1/2}\dot{M}_*^{1/2} \nonumber & \\
\ &\ \ \ \ \times x^{-2}\mathcal{C}^{-1/2}\mathcal{R}^{-1/4}\mathcal{P}^{1/2}\ll 1\,, &
\end{flalign}
it is
\begin{flalign} \label{equation:nout}
n_0({\rm H})_{\rm out}&=(3.06\times10^{25}\,{\rm cm^{-3}})\,\alpha^{-7/10}M_*^{-5/4}\dot{M}_*^{11/20} \nonumber & \\
\ &\ \ \ \ \times x^{-43/10}\mathcal{C}^{3/20}\mathcal{D}^{-7/10}\mathcal{R}^{17/40}\mathcal{P}^{11/20}\,. &
\end{flalign}
In equations (\ref{equation:conditioninn})--(\ref{equation:nout})
we have taken $M_*=M/3M_{\odot}$, $\dot{M}_*$ is the mass accretion 
rate in units of $10^{17}$ g s$^{-1}$ and $x=(r/r_{\rm g})^{1/2}$.
We refer the reader to \citet{co17} for the complete expressions 
of the functions $\mathcal{C}$, $\mathcal{D}$, $\mathcal{R}$ 
and $\mathcal{P}$. Transitions among the different regions
are determined by checking the validity of conditions 
(\ref{equation:conditioninn}), (\ref{equation:conditionmid})
and (\ref{equation:conditionout}).

In order to obtain the hydrogen density $n({\rm H})$, calculated
at the altitude $z_*$ over the equatorial plane where returning 
photons are eventually absorbed by the disk material, we need 
to specify the vertical structure of the disk. For the sake of 
simplicity, and since a full treatment of the disk vertical 
structure is beyond the scope of this work, we adopted a simple 
gaussian prescription for the density \cite[see][]{ss73},
\begin{flalign} \label{equation:verticaldensity}
n(\rm H)&=n_0({\rm H})\exp\left(-\frac{z^2}{h^2}\right)\,, &
\end{flalign}
where $h$ denotes the typical disk height at the distance $r$
from the center. In the three above-mentioned regions it turns
out to be
\begin{flalign} \label{equation:h}
h_{\rm inn}&=(0.5\,r_{\rm g})M_*^{-1}\,\dot{M}_* \nonumber & \\
\ &\ \ \ \ \times x^{-3}\mathcal{C}^{-1}\mathcal{R}^{-1}\mathcal{P} & \nonumber \\
h_{\rm mid}&=(7.0\times10^{-3}\,r_{\rm g})\,\alpha^{-1/10}M_*^{-3/10}\dot{M}_*^{1/5} \nonumber & \\
\ &\ \ \ \ \times x^{-1/10}\mathcal{C}^{-1/10}\mathcal{D}^{-1/10}\mathcal{R}^{-1/2}\mathcal{P}^{1/5} \nonumber & \\
h_{\rm out}&=(3.8\times10^{-3}\,r_{\rm g})\,\alpha^{-1/10}M_*^{-1/4}\dot{M}_*^{3/20} & \nonumber \\
\ &\ \ \ \ \times x^{1/10}\mathcal{C}^{-1/20}\mathcal{D}^{-1/10}\mathcal{R}^{-19/40}\mathcal{P}^{3/20}\,. &
\end{flalign}
A more general profile for the vertical structure \cite[still
within the assumption of geometrically thin disks, see e.g.][]{dh06}
will be addressed in future investigations.

We then resorted to choose $z_*$ in such a way that the scattering
optical depth calculated up to infinity is equal to $1$, i.e.
\begin{flalign} \label{equation:zstaropticaldepth}
\tau&=\int_{z_*}^\infty n_0({\rm H})\exp\left(-\frac{z^2}{h^2}\right)\sigma_{\rm T}\der z\equiv1\,, &
\end{flalign}
with $\sigma_{\rm T}$ the Thomson cross section. Solving numerically
equation (\ref{equation:zstaropticaldepth}) one obtains
\begin{flalign} \label{equation:zstar}
z_*(r)&=h(r)\,\erf^{-1}\left[1-\frac{2}{n_0({\rm H},r)h(r)\sigma_{\rm T}\sqrt{\pi}}\right]\,, &
\end{flalign}
where $\erf^{-1}$ denotes the inverse of the error function.

In all the expressions above, free parameters are the BH mass $M$
and accretion rate $\dot{M}$, the parameter $\alpha$ and the height
$h_0$ of the disk at the innermost stable circular orbit. Throughout 
our calculations we take $\alpha=0.2$ \cite[see][]{co17} and
\begin{flalign} \label{equation:h0}
h_0^{\rm inn}&=0.002\,r_{\rm g} \nonumber & \\
h_0^{\rm mid}&=(1.8\times10^{-3}\,r_{\rm g})\,\alpha^{1/8}M_*^{-3/8}\dot{M}_*^{1/4} \nonumber & \\
\ &\ \ \ \ \times x_0^{1/8}\mathcal{C}_0^{-1/8}\mathcal{R}_0^{-1/2} \nonumber & \\
h_0^{\rm out}&=(1.3\times10^{-3}\,r_{\rm g})\,\alpha^{1/17}M_*^{-5/17}\dot{M}_*^{3/17} \nonumber & \\
\ &\ \ \ \ \times x_0^{5/17}\mathcal{C}_0^{-1/17}\mathcal{D}_0^{-1/34}\mathcal{R}_0^{-8/17}\,, &
\end{flalign}
where a $0$ subscript denotes quantities calculated at $r=\rms$. The
chosen values of $M$ and $\dot{M}$ are specified in the text.

\label{lastpage}

\end{document}